\documentclass{aa}  

\usepackage{graphicx}
\usepackage{txfonts}
\usepackage[colorlinks=true,citecolor=blue]{hyperref}
\usepackage{siunitx}

\usepackage{orcidlink}

\newcommand\grizli{\ensuremath{\textsc{grizli}}}
\newcommand\pygcg{\ensuremath{\textsc{pygcg}}}
\newcommand\pygrife{\ensuremath{\textsc{pygrife}}}
\newcommand\zorig{\ensuremath{z_{\rm{orig}}}}
\newcommand\zphot{\ensuremath{z_{\rm{phot}}}}
\newcommand\zspec{\ensuremath{z_{\rm{spec}}}}
\newcommand\zniriss{\ensuremath{z_{\rm{NIRISS}}}}
\newcommand\zgrizli{\ensuremath{z_{\rm{\grizli}}}}
\newcommand\flambda{\ensuremath{f_{\lambda}}}
\newcommand{\e}[1]{\ensuremath{\times 10 ^{#1}}}
\newcommand{\Oiii}{\ensuremath{\left[\ion{O}{iii}\right]}}
\newcommand{\Oii}{\ensuremath{\left[\ion{O}{ii}\right]}}
\newcommand{\Oiilam}{\ensuremath{\left[\ion{O}{ii}\right]\,{\lambda 3727,3729}}}
\newcommand{\Nii}{\ensuremath{\left[\ion{N}{ii}\right]}}
\newcommand{\Sii}{\ensuremath{\left[\ion{S}{ii}\right]}}
\newcommand{\Siii}{\ensuremath{\left[\ion{S}{iii}\right]}}
\newcommand{\Halpha}{\ensuremath{\ion{H}{$\alpha$}}}
\newcommand{\Hbeta}{\ensuremath{\ion{H}{$\beta$}}}
\newcommand{\Hgamma}{\ensuremath{\ion{H}{$\gamma$}}}

\newcommand{\PaB}{\ensuremath{\ion{Pa}{$\beta$}}}

\begin{document} 

\title{The GLASS-JWST Early Release Science programme: The NIRISS spectroscopic catalogue}

\author{
Peter J. Watson
\inst{1}\orcidlink{0000-0003-3108-0624}
\and
Benedetta Vulcani
\inst{1}\orcidlink{0000-0003-0980-1499}
\and
Tommaso Treu
\inst{2}\orcidlink{0000-0002-8460-0390}
\and
Guido Roberts-Borsani
\inst{3}\orcidlink{0000-0002-4140-1367}
\and
Nicol\`o Dalmasso
\inst{4,5}\orcidlink{0000-0002-1850-4050}
\and
Xianlong He
\inst{2,6,7}\orcidlink{0000-0002-1336-5100}
\and
Matthew A. Malkan
\inst{2}\orcidlink{0000-0001-6919-1237}
\and
Takahiro Morishita
\inst{8}\orcidlink{0000-0002-8512-1404}
\and
Sofía Rojas Ruiz
\inst{2}\orcidlink{0000-0003-2349-9310}
\and
Yechi Zhang
\inst{8}\orcidlink{0000-0003-3817-8739}
\and
Ayan Acharyya
\inst{1}\orcidlink{0000-0003-4804-7142}
\and
Pietro Bergamini
\inst{9,10}\orcidlink{0000-0003-1383-9414}
\and
Maru\v{s}a Brada\v{c}
\inst{11}\orcidlink{0000-0001-5984-0395}
\and
Adriano Fontana
\inst{12}\orcidlink{0000-0003-3820-2823}
\and
Claudio Grillo
\inst{9,13}\orcidlink{0000-0002-5926-7143}
\and
Tucker Jones
\inst{14}\orcidlink{0000-0001-5860-3419}
\and 
Danilo Marchesini
\inst{15}\orcidlink{0000-0001-9002-3502}
\and
Themiya Nanayakkara
\inst{16}\orcidlink{0000-0003-2804-0648}
\and
Laura Pentericci
\inst{12}\orcidlink{0000-0001-8940-6768}
\and
Chanita Tubthong
\inst{15}\orcidlink{0000-0002-7907-2634}
\and
Xin Wang
\inst{17,18,19}\orcidlink{0000-0002-9373-3865}
}

\institute{
INAF -- Osservatorio Astronomico di Padova, Vicolo Osservatorio 5, 35122 Padova, Italy
\email{peter.watson@inaf.it}
\and  
University of California, Los Angeles, Department of Physics and Astronomy, 430 Portola Plaza, Los Angeles, CA 90095, USA
\and  
Department of Physics and Astronomy, University College London, Gower Street, London WC1E 6BT, UK
\and  
School of Physics, University of Melbourne, Parkville, Vic 3010, Australia
\and
Australian Research Council Centre of Excellence for All-Sky Astrophysics in 3-Dimensions, Australia
\and
School of Astronomy and Space Science, University of Chinese Academy of Sciences (UCAS), Beijing 100049, People's Republic of China
\and
School of Physics and Technology, Wuhan University (WHU), Wuhan 430072, People's Republic of China
\and
IPAC, California Institute of Technology, 1200 E. California Blvd, Pasadena, CA 91125, USA
\and 
Dipartimento di Fisica, Università degli Studi di Milano, Via Celoria 16, I-20133 Milano, Italy
\and
INAF – OAS, Osservatorio di Astrofisica e Scienza dello Spazio di Bologna, via Gobetti 93/3, I-40129 Bologna, Italy
\and
University of Ljubljana, Department of Mathematics and Physics, Jadranska ulica 19, SI-1000 Ljubljana, Slovenia
\and
INAF – Osservatorio Astronomico di Roma, via Frascati 33, 00078
Monteporzio Catone, Italy
\and
INAF – IASF Milano, via A. Corti 12, I-20133 Milano, Italy
\and
Department of Physics and Astronomy, University of California, Davis, 1 Shields Ave, Davis, CA 95616, USA
\and
Department of Physics and Astronomy, Tufts University, 574 Boston Ave., Medford, MA 02155, USA
\and
Centre for Astrophysics and Supercomputing, Swinburne University of Technology, PO Box 218, Hawthorn, VIC 3122, Australia
\and
School of Astronomy and Space Science, University of Chinese Academy of Sciences (UCAS), Beijing 100049, China
\and
National Astronomical Observatories, Chinese Academy of Sciences, Beijing 100101, China
\and
Institute for Frontiers in Astronomy and Astrophysics, Beijing Normal University,  Beijing 102206, China
}

\authorrunning{Peter J. Watson et al.}
\titlerunning{GLASS-JWST ERS: The NIRISS spectroscopic catalogue}

\date{Received 1 April, 2025; accepted 28 May, 2025}

\abstract
{
We present a spectroscopic redshift catalogue of sources within the Abell 2744 cluster field, derived from JWST/NIRISS observations, obtained as part of the GLASS-JWST Early Release Science programme. We describe the data reduction, the contamination modelling, and the source detection, as well as the data quality assessment, the redshift determination, and the validation. 
The catalogue consists of 354 secure and 134 tentative redshifts, of which 245 are new spectroscopic redshifts, spanning the range $0.1\leq z \leq 8.2$.
These include 17 galaxies at the cluster redshift, one galaxy at $z\approx8$, and a triply imaged galaxy at $z=2.653\pm0.002$.
Comparing against galaxies with existing spectroscopic redshifts (\zspec), we find a small offset of $\Delta z = (\zspec-\zniriss)/(1+\zspec)=(1.3\pm1.6)\e{-3}$.
We also present a forced extraction tool (\pygrife) and a visualisation tool (\pygcg) to the community, to aid with the reduction and classification of grism data. 
This catalogue will enable future studies of the spatially resolved properties of galaxies throughout cosmic noon, including dust attenuation and star formation.
As a first application of the catalogue, we discuss the spectroscopic confirmation of multiple image systems and the identification of multiple overdensities at $1<z<2.7$.
}


   \maketitle
%
%-------------------------------------------------------------------

\nolinenumbers
\section{Introduction} \label{sec:introduction}

Observations of galaxy clusters are essential to our understanding of galaxy formation and evolution. Galaxies within clusters are ideal for studying the environmental processes that can regulate star formation, whilst the most massive clusters magnify background galaxies through strong gravitational lensing, greatly aiding studies of the distant universe. Essential to all of these scientific cases, however, is our ability to derive accurate and reliable redshifts for the galaxies involved. The redshifts of cluster members can be used to estimate the mass distribution of the cluster \citep[e.g.][]{wang_grism_2015,jauzac_extraordinary_2016,bergamini_glass-jwst_2023,furtak_uncovering_2023}, and obtaining spectroscopic redshifts of multiply imaged background galaxies is vital for constraining strong-lensing models \citep[e.g.][]{bergamini_new_2023,rihtarsic_canucs_2025}. In turn, these lens models can be used to derive the magnification factor of background galaxies and, hence, their intrinsic properties.

However, obtaining a large sample of spectroscopic redshifts can be a costly endeavour. Multi-object spectroscopy (MOS), whether slit- or fibre-based, relies on prior imaging in order to choose targets, introducing a strong selection bias, and often faces instrumental constraints on the density of targets in a single observation. Integral-field spectroscopy (IFS) with instruments such as the Multi Unit Spectroscopic Explorer \citep[MUSE;][]{bacon_muse_2010} has no such restrictions on target selection but instead typically covers a smaller field of view ($\lesssim1\,\rm{arcmin}^2$). An alternative method is slitless spectroscopy, whereby the light from every object in the field is dispersed onto the detector. Whilst first used in ground-based observations in 1903 at the Lick Observatory \citep{palmer_application_1903}, its use saw a resurgence with the launch of the \textit{Hubble} Space Telescope (HST), which no longer had to contend with telluric contamination, enabling efficient, wide-field spectroscopic surveys \citep[e.g.][]{mccarthy_emission-line_1999,atek_wfc3_2010,brammer_3d-hst_2012,treu_grism_2015,pirzkal_figsfaint_2017}.

This has continued with the launch of the \textit{James Webb} Space Telescope (JWST), which carries the Near Infrared Imager and Slitless Spectrograph \citep[NIRISS;][]{doyon_jwst_2012} and the Near Infrared Camera \citep[NIRCam;][]{rieke_overview_2005}, both of which permit observations in wide-field slitless spectroscopy (WFSS) mode \citep{greene__2017,willott_near-infrared_2022}. The wavelength coverage of NIRISS WFSS ($\approx0.8$-2.2$\,\mu$m) and NIRCam WFSS ($\approx2.4$-5.0$\,\mu$m)\footnote{These wavelength ranges are the maximum extent covered by each instrument and require observations taken through multiple filters.} provide the perfect complement to ground-based optical spectrographs such as MUSE, which are most effective at obtaining redshifts in the ranges $z\lesssim 1.5$ and $3\lesssim z\lesssim6$ \citep{richard_atlas_2021,bacon_muse_2023}. The WFSS mode of NIRISS has seen particular use for pure parallel surveys \citep{malkan_passage-parallel_2021,glazebrook_jwst_2023} and for targeting dense cluster environments \citep{willott_near-infrared_2022,noirot_first_2023}.

\begin{figure} \label{fig:pointing_location}
    \centering
    \includegraphics[width=\columnwidth]{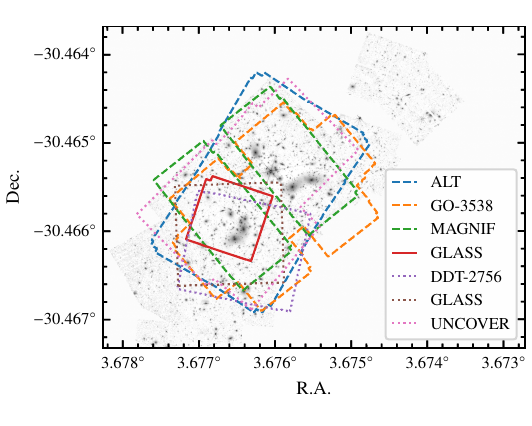}
    \caption{Relative coverage of all JWST spectroscopic observations in the A2744 field. We delineate the NIRCam WFSS observations by dashed lines and the NIRSpec pointings by dotted lines. The NIRISS observations used in this analysis are indicated by the solid line. In the background, we show the field as seen through the F200W filter on NIRCam, taken from the MegaScience combined mosaics \citep{suess_medium_2024}.}
\end{figure}

In this paper, we present a redshift catalogue derived from NIRISS WFSS observations of the Abell 2744 galaxy cluster, also known as `Pandora's Cluster' \citep{abell_catalog_1989,merten_creation_2011}. Abell 2744, hereinafter A2744 ($\rm{R.A.}=3.58641$, $\rm{Dec.}=-30.39997$; $\zspec=0.3064$), is amongst the most-observed strong-lensing galaxy clusters, and we show a wide-field view in Fig.~\ref{fig:pointing_location}. It was observed using HST WFSS in the Grism Lens-Amplified Survey from Space \citep[GLASS;][]{treu_grism_2015}, selected as part of the Hubble Frontier Fields initiative \citep[HFF;][]{lotz_frontier_2017}, and followed up by the MUSE Guaranteed Time Observing (GTO) programme \citep{mahler_strong-lensing_2018,richard_atlas_2021}. It has a wealth of ancillary data spanning the entire electromagnetic spectrum, from X-ray \citep{kempner_chandra_2004,owers_dissection_2011} to radio continuum \citep{pearce_vla_2017,rajpurohit_dissecting_2021}. More recently, A2744 has been the focus of many complementary JWST programmes, including GLASS-JWST Early Release Science \citep[GLASS-JWST ERS;][]{treu_glass-jwst_2022}, UNCOVER \citep{bezanson_jwst_2024}, and MegaScience \citep{suess_medium_2024}, enabling a complete coverage in all NIRCam photometric bands from 0.6-5$\,\mu$m. It has also been targeted using the MOS mode of NIRSpec \citep{mascia_glass-jwst_2024,bezanson_jwst_2024} and NIRCam WFSS \citep{iani_unveiling_2023,sun_magnif_2023,naidu_all_2024}.
The NIRISS data presented here have already enabled many specific science cases, such as the identification of Lyman-break galaxies at $z\geq7$ \citep{roberts-borsani_early_2022}, protoglobular clusters at $z=4$ \citep{vanzella_early_2022}, low-mass quiescent galaxies at $z\geq2$ \citep{marchesini_early_2023}, a magnified star at $z=2.65$ \citep{chen_early_2022}, measurements of metallicity at $z=2-3$ \citep{wang_early_2022,li_mass-metallicity_2023,he_early_2024}, and extreme \Oiii\ equivalent widths between $z=1-3.4$ \citep{boyett_early_2022}.

This paper is structured as follows. In Sect.~\ref{sec:data} we describe the data, including the observing strategy and in Sect.~\ref{sec:method} the methods used to reduce and process them. In Sect.~\ref{sec:results} we present the final catalogue, validating against previous observations in the field and identifying a select few sources for which our redshifts differ from those in the existing literature. We discuss potential multiply imaged systems and background overdensities in the full redshift distribution before summarising our work in Sect.~\ref{sec:conclusions}. Throughout this paper, coordinates are given in the International Celestial Reference System (ICRS); magnitudes are in the AB system; and we assume a standard $\Lambda$ cold dark matter cosmology, with $\Omega_{\rm{M}}=0.3$, $\Omega_{\Lambda}=0.7$, and $h=0.7$.

\section{Data} \label{sec:data}

\subsection{Instrumental characteristics} \label{sec:instrument}

The data were taken with JWST/NIRISS in WFSS mode \citep{doyon_jwst_2012,willott_near-infrared_2022}. The NIRISS detector consists of a $2048 \times 2048$ pixel Hawaii-2RG sensor. The pixel size is $18\times18\,\mu$m, leading to an average scale of $0\farcs0653\times0\farcs0658/\rm{pixel}$ across a $\ang[angle-symbol-over-decimal]{;2.2;}\times\ang[angle-symbol-over-decimal]{;2.2;}$ field of view, varying slightly with both the position on the detector and the optical elements chosen. Within the optical path of the instrument, there exist user-selectable elements in both a pupil and filter wheel. For WFSS, the filter wheel contains a ‘CLEAR’ filter, which allows for direct imaging, and two identical grisms ‘GR150R’ and ‘GR150C’, mounted so that their dispersion directions are perpendicular to each other on the detector, and aligned with the rows and columns of the detector, respectively. The wavelength dispersion of both grisms has a small dependence on the detector position, but is linear for the first order at $\approx47\textup{\AA}/\rm{pixel}$ \citep{pacifici_niriss_2022}. The spatial point-spread function of NIRISS varies slowly with wavelength, from 0\farcs088 at $0.9\,\mu$m to 0\farcs11 at $2.2\,\mu$m, and is undersampled by the detector. For a point source, the spectral resolving power is therefore defined by the wavelength interval over 2 pixels ($\Delta\lambda\approx94\,\textup{\AA}$ at the field centre), giving $R\equiv\lambda/\Delta\lambda\approx150$ at 1.4\,$\mu$m. The NIRISS pupil wheel contains broad- and medium-band blocking filters, defining the wavelength coverage. In WFSS mode, these are either combined with the clear filter to obtain direct images of the field, or with the grisms to obtain dispersed spectra over a particular wavelength range.

\subsection{Observing strategy} \label{sec:observations}

\begin{figure} 
    \centering
    \includegraphics[width=\columnwidth]{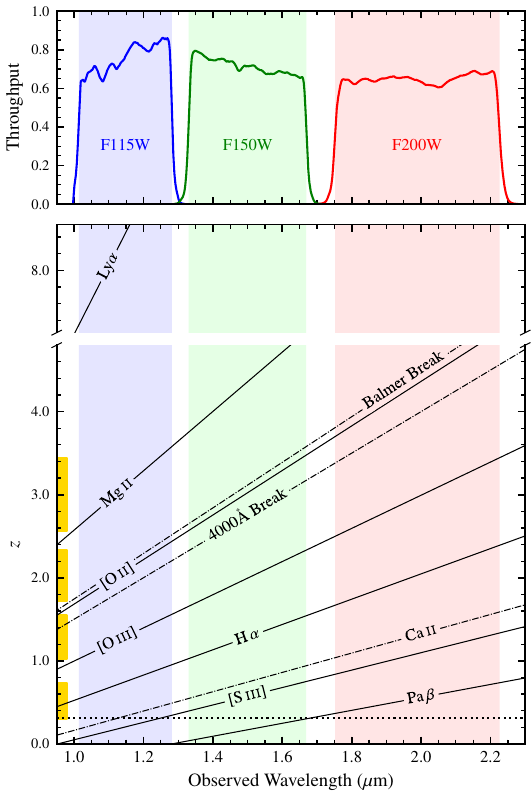}
    \caption{Top: Throughput of the three broad-band NIRISS filters used in this analysis: F115W, F150W, and F200W.
    The shaded regions correspond to the wavelength range within which the throughput is greater than 50\% of the peak value.
    Bottom: Visibility of a selection of strong optical-NIR emission lines (solid lines) and absorption features (dot-dashed lines), varying with redshift.
    On the left of the plot, we show the approximate redshift ranges in which multiple strong lines are visible.
    The cluster redshift ($z=0.3064$) is denoted by the dashed line. }
    \label{fig:wavelength_coverage}
\end{figure}

The JWST/NIRISS data were taken as part of the GLASS-JWST ERS programme \citep{treu_glass-jwst_2022,roberts-borsani_early_2022}.
These consist of a single pointing in the Abell 2744 cluster field, shown in Fig.~\ref{fig:pointing_location} in reference to other JWST spectroscopic observations, covering a $\ang[angle-symbol-over-decimal]{;2.2;}\times\ang[angle-symbol-over-decimal]{;2.2;}$ field of view.
The programme design called for observations in three wide-band blocking filters (F115W, F150W, and F200W), spanning a range of approximately 1.0-2.2 $\mu$m, with the filter transmission curves shown in the upper panel of Fig.~\ref{fig:wavelength_coverage}.
The observations were executed on 28-29 June 2022 and 7 July 2023, resulting in $\approx18$\,hours of WFSS.
The observations were taken using a four-point dither pattern with offsets of $\approx0\farcs6$, and split between the ‘CLEAR’ (direct imaging), ‘GR150R’, and ‘GR150C’ elements in the NIRISS filter wheel.
The total exposure times were 5497s for direct imaging and 10307s for each of the GR150R and GR150C grisms, in the F115W and F150W filters, and half that for the F200W filter.
For these observations, the detector was read out in the ‘NIS’ pattern, using the full frame readout mode.

The use of both grisms to provide two orthogonal dispersion directions was an essential component of these observations.
In slitless spectroscopy, the light from all objects is dispersed in one direction, resulting in the blending of the spectra of nearby objects along the dispersion axis.
This also creates a potentially severe degeneracy between the on-sky position and wavelength along the spectral axis for individual sources.
This degeneracy can be minimised, if not broken entirely, by obtaining dispersed spectra at multiple orientations \citep{pirzkal_two-dimensional_2018}.
Two orientations, separated by \ang{90;;}, allow for an efficient use of observing time, whilst retaining the ability to de-blend almost all sources in the field, and had previously been used to great effect with HST in the GLASS survey \citep{schmidt_through_2014,treu_grism_2015,vulcani_grism_2015,vulcani_grism_2016}.

\section{Method} \label{sec:method}

\subsection{Data reduction} \label{sec:data_reduction}

The data were downloaded from the Mikulski Archive for Space Telescopes (MAST)\footnote{\href{https://mast.stsci.edu/search/ui/\#/jwst}{archive.stsci.edu}}, in the form of uncalibrated ramp exposures (\texttt{*\_uncal.fits}).
These were reduced to count-rate files (\texttt{*\_rate.fits}) using the \textsc{Detector1} module from the official STScI calibration pipeline\footnote{\href{https://github.com/spacetelescope/jwst}{github.com/spacetelescope/jwst}}.
This performed a number of detector-level calibrations, such as persistence correction, dark current and bias subtraction, and slope fitting.
The operational context used was ‘jwst\_1173.pmap’, and the grism trace configuration was \texttt{*221215.conf} \citep{matharu_updated_2022}.
Although more recent trace configurations were available \citep{pirzkal_next_2024}, these showed significant regressions in the modelling and subtraction of the dispersed $0^{\rm{th}}$ orders, which were deemed to be vital for the reduction of such a crowded field.

The remainder of the data reduction used the Grism Redshift \& Line package \citep[\grizli;][]{brammer_grizli_2019}.
We used \grizli's \textsc{aws.visit\_processor} pipeline to perform flat-fielding, $1/f$ correction, sky background subtraction, and astrometric alignment for all observations, both imaging and WFSS.
In general, we followed the default \grizli\ \textsc{visit\_processor} parameters (as of version 1.12.8).
Observations were aligned to the astrometric reference frame of the Legacy Surveys DR9 \citep{dey_overview_2019}.
We used \grizli\ to produce drizzled mosaics from the direct imaging, at a scale of $\ang[angle-symbol-over-decimal]{;;0.03}/\rm{pixel}$ (compared to the native detector pixel scale of $\ang[angle-symbol-over-decimal]{;;0.066}/\rm{pixel}$), using a square kernel with $\textsc{pixfrac}=0.8$, the ratio of the linear size of the drizzled pixels to the input.

\subsection{Contamination modelling} \label{sec:contamination_modelling}

One of the challenges of slitless spectroscopic data is removing the dispersed contaminating flux from nearby objects.
While there exist alternative methods \citep{bushouse_jwst_2025}, here we used the standard procedures implemented in \textsc{grizli}.
For each detected source in the direct imaging, we forward modelled the dispersed flux for all combinations of filters and grism orientations, assuming a flat $f_{\lambda}$ spectrum.
This created a so-called ‘contamination map’, enabling (to first order) the removal of overlapping spectra from the object of interest.
As in \cite{roberts-borsani_early_2022}, all pixels used for extraction are weighted by $$w_i=\sigma^{-2}_i \cdot\exp\left(-a\cdot\lvert f_{{\rm{cont}},i}\rvert/\sigma_i\right),$$ where for any pixel $i$, $\sigma_i$ is the flux uncertainty, $f_{{\rm{cont}},i}$ is the flux in the contamination model, and $a$ is a constant factor.
Here, we adopted the factor $a=0.2$ for down-weighting contaminated pixels.

\subsection{Source detection and de-blending} \label{sec:source_detection}

\begin{figure*}
    \centering
    \includegraphics[width=\textwidth]{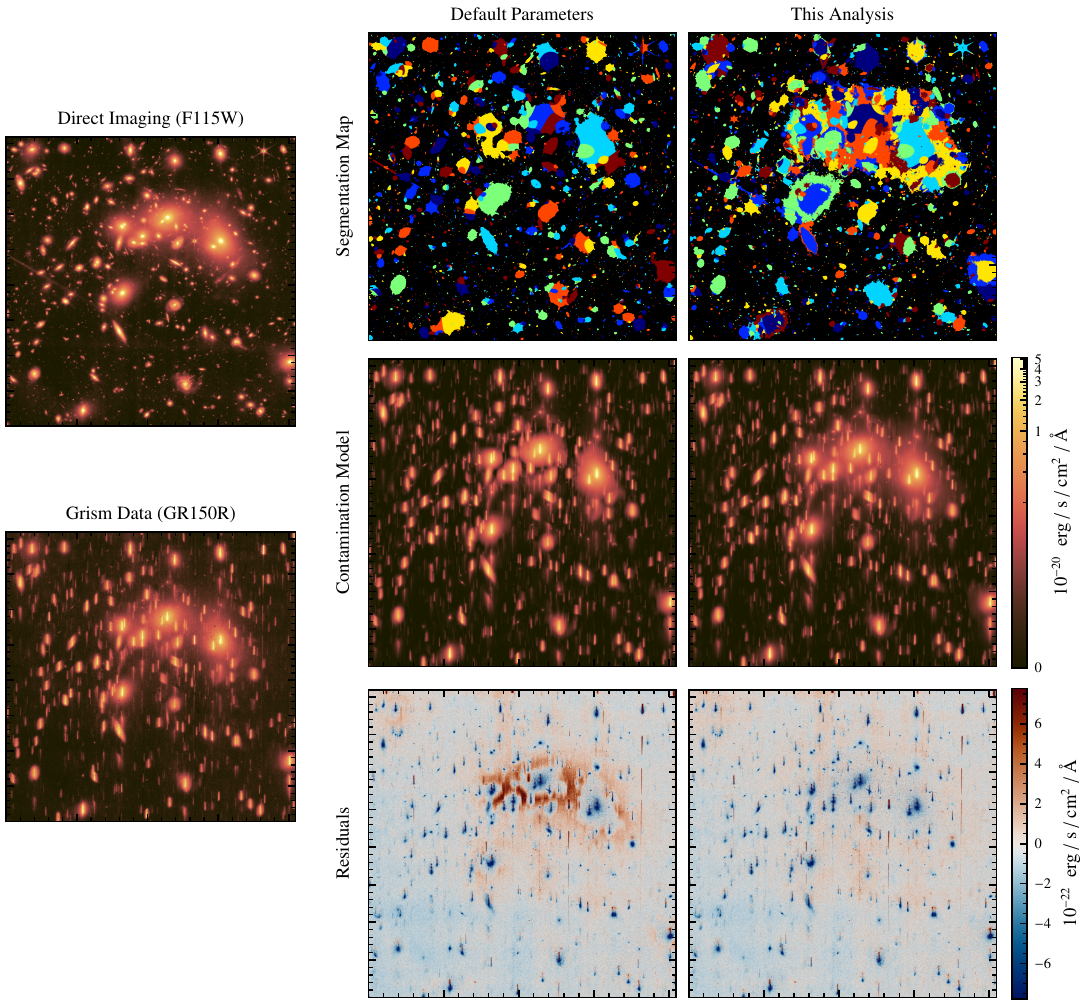}
    \caption{Left: Examples of the observed data, specifically the drizzled direct imaging in the F115W blocking filter and the corresponding dispersed spectra using the GR150R grism.
    The colour map for these plots uses a logarithmic scale, with the colour bar shown on the right.
    Top right: Derived segmentation map for both the default \grizli\ parameters (left) and the parameters used in this analysis (right), following all inspection and additional de-blending steps.
    Middle right: Contamination model derived using each of the segmentation maps.
    Note that the default reduction pipeline isolates many of the extended galaxies from the diffuse cluster light.
    Bottom-right: Residual flux after subtracting each contamination model from the observations.
    The colour map here adopts a linear scale and covers a much narrower range.
    }
    \label{fig:contamination_model_comparison}
\end{figure*}

To identify sources in the field of view, we produced a flux-weighted stacked image from the three individual filter mosaics.
However, we wish to stress that the default parameters for background subtraction and source detection are not suitable for a crowded cluster field such as A2744.
Typically, one determines the background (from both the sky and the detector) by dividing the image into a low-resolution grid, and using techniques such as sigma-clipping to derive local estimates.
This requires careful tuning of the grid mesh size, in order to encompass the largest sources in the image, but preserve small-scale variations in the background level.
However, an additional complication here is that the field is affected by the extended intracluster light (ICL), and a simple background estimation cannot adequately distinguish between this and the true background level.
Although subtracting this diffuse component showed improvements in the detection and de-blending of faint, compact sources, this resulted in a sparse segmentation map that did not fully cover the flux in the field, as shown in Fig.~\ref{fig:contamination_model_comparison}.
Following the reduction process, we obtained a contamination model with unphysical discontinuities, arising from the ICL contribution to the outer regions of cluster galaxies not being modelled.
These discontinuities affect individual object extractions, where they can be confused with Lyman or Balmer breaks, and reduce the S/N.

To counteract this, we performed a two-step extraction procedure. The initial source detection step turned off the background subtraction in \textsc{sep} (the \textsc{python} port of Source Extractor, \citealt{bertin_sextractor_1996}), generating a segmentation map with all objects detected at 1$\sigma$ above the background, after filtering with a matched convolution kernel. At this stage, 3493 objects were detected, and the segmentation map was used to create a contamination model, where we initially assumed a flat \flambda\ spectrum for all sources. We then used \grizli\ to fit a seventh-order polynomial to the spectrum of each source, refining this contamination model over three iterations. Whilst there are alternative methods to account for the diffuse ICL (see \citealt{estrada-carpenter_when_2024} for one such procedure), we find that this simple change reduces the residual flux from the contamination model by approximately $40\%$.

During the first stage of the visual inspection (see Sect.~\ref{sec:inspection_first_stage}), sources were flagged if their segmentation map contained multiple distinct objects (and thus required further de-blending), or was overly extended due to the faint ICL. For these sources, we generated additional segmentation maps, using both the default background subtraction and the more aggressive de-blending parameters. In a select few cases, further manual adjustments were necessary to ensure an optimal separation between overlapping sources, including eight objects, which were separated into multiple components in the initial segmentation map. Using these new segmentation maps, we re-extracted only the new sources, using the existing complete contamination model, via a forced extraction tool developed for this exact purpose, \pygrife\footnote{\href{https://pygrife.readthedocs.io/}{pygrife.readthedocs.io/}}. This tool enabled a fast comparison of the optimal detection and de-blending parameters required for each source, without needing to recreate the full contamination map on every iteration, and is described in more detail in Appendix~\ref{app:pygrife}. This resulted in a final detection catalogue containing 3652 sources.

We show in Fig.~\ref{fig:contamination_model_comparison} a comparison between the default source detection parameters used in \grizli, and those used in this analysis. The adopted segmentation map covers $\approx40\%$ more of the direct image. The differences in the contamination model are clearly visible, with the default pipeline only covering the inner portion of the largest galaxies, and almost none of the ICL. This results in a large fraction of unsubtracted flux in the residual image, shown at the bottom of Fig.~\ref{fig:contamination_model_comparison}. We note that the two methods lead to some oversubtractions in the centre of the brightest galaxies. This is due to the spatial variation of the stellar populations within these galaxies, which cannot be modelled by \grizli\ at present (although see \citealt{estrada-carpenter_when_2024} for an alternative solution). Similarly, there are a number of bright streaks remaining in the residual image, typically from saturated sources in the direct imaging. No solution as yet exists to model the contribution from these sources, and extracted spectra overlapping with these areas were typically discarded in our analysis.

\subsection{Redshift priors} \label{sec:redshift_priors}

One of the most computationally expensive components of the data reduction is the extraction and subsequent redshift fitting of all individual objects. An unconstrained redshift search using only grism data may also result in degenerate solutions, particularly if the spectra have low S/N, or only a single line is detected above the noise floor. For this analysis, we therefore utilised existing ancillary data to provide informative priors. We matched our source detection catalogue (see Sect.~\ref{sec:source_detection}) against the ALT spectroscopic catalogue, derived from NIRCam WFSS observations \citep{naidu_all_2024}, and the spectroscopic compilation presented by \citet{merlin_astrodeep-jwst_2024}. The latter included spectroscopic redshifts sourced from NIRSpec, HST, MUSE, and other ground-based instruments \citep{couch_spectroscopic_1987,couch_morphological_1998,boschin_internal_2006,braglia_multi-wavelength_2009,owers_dissection_2011,schmidt_through_2014,treu_grism_2015,bergamini_glass-jwst_2023,mascia_glass-jwst_2024}. We also included the most recent NIRSpec-derived redshifts from UNCOVER DR4 \citep{price_uncover_2025}. For objects without existing spectroscopic redshifts, we obtained photometric redshifts from the ASTRODEEP catalogue \citep{merlin_astrodeep-jwst_2024}.

In the case where an object in our catalogue had an existing spectroscopic redshift (\zspec), we restricted the redshift fit to search in the range $\zspec\pm1\%$, or $\zspec\pm0.01$, whichever resulted in the wider range. For objects which have only a photometric redshift (\zphot), we searched in the range $\zphot\pm10\%$ initially. For any objects for which no match was found in either the spectroscopic or photometric catalogues, we searched over the range $0\leq z \leq 10$.

\subsection{Object extraction} \label{sec:object_extraction}

\begin{figure*} 
    \centering
    \includegraphics[width=\textwidth]{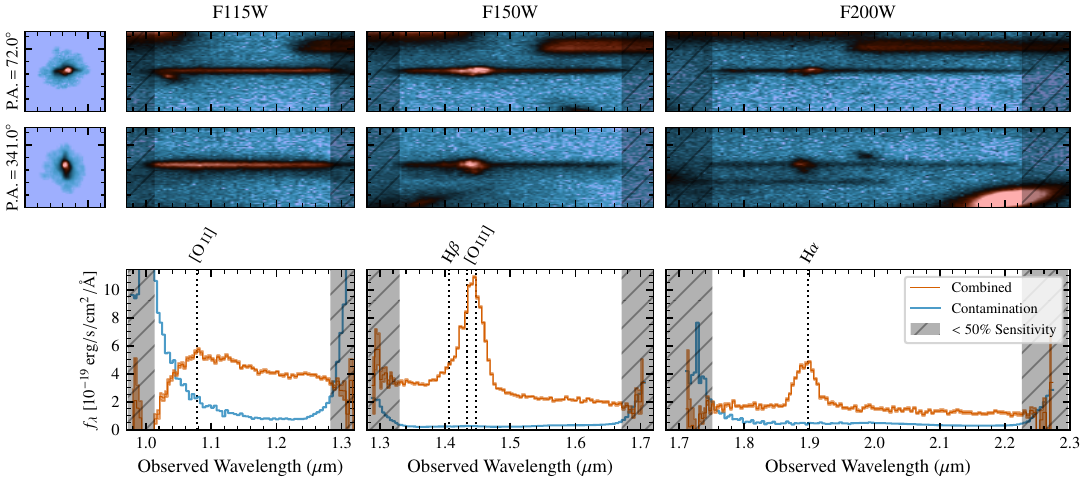}
    \caption{Example galaxy (ID: 2938, $z=1.891$). On the left, we show direct image cut-outs of the galaxy in the F200W filter, noting that equivalent cut-outs are also available for the F115W and F150W filters. Upper row: Stacked 2D dispersed spectrum for the galaxy using the GR150R grism, equivalent to dispersing the light along a position angle of \ang{72}. Off-centre contamination from other nearby sources is clearly visible. Middle row: As above, but for the GR150C grism, equivalent to a position angle of \ang{341}. The asymmetric morphology results in a clear difference in the shape of the dispersed spectrum, most noticeably around the emission lines. Bottom row: Extracted 1D spectra from a stack of the two grism orientations. We show also the 1D equivalent spectrum from the contamination in each filter and mark the position of several emission lines. In all panels, the grey shaded regions denote the wavelength range where the filter sensitivity drops to less than 50\% of the peak value.}
    \label{fig:niriss_beams_example}
\end{figure*}

For each detected source, we extracted the 2D spectra and searched for the best-fit redshift using \grizli. We modelled all individual grism exposures simultaneously, preserving the collective S/N whilst avoiding complications associated with stacking (including, but not limited to, morphological broadening, cross-dispersion offsets, and differing dispersion directions). We used the default \grizli\ template set, which includes galaxy templates generated using the flexible stellar population synthesis code of \cite{conroy_metal-rich_2018}, and a dedicated set of individual emission line templates, which allow for varying line ratios. We then forward modelled the dispersed flux for each of the template spectra, in all grism exposures, and use non-negative least squares to find the best-fitting combination of templates. This step was repeated for a range of redshifts, as detailed in Sect.~\ref{sec:redshift_priors}, and \zgrizli\ was determined as the redshift with the minimum $\chi^2$. The 1D spectrum was then optimally extracted following \cite{horne_optimal_1986} with the \textsc{optimal\_extract()} method of \grizli, using the position and morphology of the source in the direct image as a reference. An example is shown in Fig.~\ref{fig:niriss_beams_example}, where we display both the 1D and 2D spectra in all filters and orientations, showing clearly how the extended morphology of the source impacts the dispersed spectrum. This fitting and extraction procedure was possible for only 3141 sources in our catalogue. The remaining 511 sources were those at the edges of the direct imaging, where the dispersed spectra fell outside of the detector coverage. The locations and assigned IDs of these sources are included in the final catalogue for completeness, but were not used for any of the subsequent analysis.

\subsection{Visual inspection} \label{sec:visual_inspection}

\subsubsection{First stage} \label{sec:inspection_first_stage}

For the initial stage of the inspection, we focussed on removing low-quality observations from the sample.
We developed a tool, \pygcg\footnote{\href{https://pypi.org/project/pyGCG/}{pypi.org/project/pyGCG/}}, which operates on a subset of the \grizli\ reduced data products, to produce a simple interactive interface for inspection and classification (see Appendix~\ref{app:pygcg} for more details).
One of the team members inspected the 2D ‘beams’ spectra for all 3141 extracted objects.
Individual beams (the stacked exposures for a particular filter and grism orientation) were flagged as ‘Unusable’ if they satisfied at least one of the following criteria:
\begin{itemize}
    \item The object is merely an artefact in the direct imaging used for source detection (e.g. uncorrected hot pixels or cosmic rays).
    \item The contamination has been oversubtracted, and no signal is visible.
    \item The contamination has been undersubtracted, and outshines the expected signal.
    \item The dispersed spectrum has been severely truncated by the edge of the detector, such that no pixels lie in the region where the filter transmission is greater than 50\% of the peak value (see Fig.~\ref{fig:wavelength_coverage}).
    \item The dispersed spectrum is substantially affected by artefacts, which can be either optical or from the data reduction.
\end{itemize}
Examples of all of these are shown in Appendix~\ref{app:examples_beam_quality_flag}.
We did not track the number of beams removed based on each of these individual criteria, as beams often satisfied multiple criteria simultaneously.

Following this stage, we discarded any object for which all beams were flagged as ‘Unusable’ (accounting for a total of 907 sources, leaving 2234 to be inspected in more detail).
In addition, we refitted any object for which one or more of the beams were flagged.
Whilst \grizli\ includes some functionality to discard heavily contaminated regions when combining observations across multiple grism orientations, we preferred to explicitly remove these data based on our initial quality assessment.

\subsubsection{Second stage} \label{sec:inspection_second_stage}

For the second stage of the inspection, we introduced a tiered system of quality flags for the \grizli\ redshift fit, as well as a procedure for assessing the quality of the stacked 2D spectra.
For the former, the flags were assigned as follows:
\begin{itemize}
    \item Secure: Multiple lines or features (e.g. the Lyman break) were visible in the extracted 1D spectrum.
    In some cases, additional lines may appear due to factors such as spatially offset star-forming clumps or unmasked contamination.
    As such, we assigned this flag only when the majority of unambiguous features were consistent with a singular redshift solution.
    Alternatively, a single strong line or feature was visible, which was consistent with the photometric or spectroscopic redshift prior for that object (see Sect.~\ref{sec:redshift_priors}).
    \item Tentative: A singular line or feature was visible, which was not attributable to contamination from other objects. This was either inconsistent with the redshift prior, or there was no existing redshift available.
    \item Undetermined: Multiple lines were visible, but were inconsistent with a singular redshift solution. 
    Alternatively, one or more features were visible in the 1D spectrum, but the most probable cause was imperfect removal of contaminating spectra, such as the dispersed $0^{\rm{th}}$ order from a brighter object.
    Objects where no features were visible also received this classification.
\end{itemize}
For the 2D spectra quality flags, we adopted the following classification:
\begin{itemize}
    \item Good: There were no significant data reduction artefacts present, and the signal from the object was free from any significant contamination.
    \item Poor: The spectrum was partially affected by artefacts. Alternatively, there was significant contamination affecting a portion ($<50\%$) of the dispersed spectrum.
    \item Unusable: The criteria here were the same as in Sect.~\ref{sec:inspection_first_stage}.
\end{itemize}
Whilst these criteria were more subjective than the redshift quality flags, we considered it highly beneficial to have an assessment of the reduced data quality. 
In particular, for studies of the spatially resolved emission, it is vital to have good quality data in the two grism orientations, and such a flag may be used to limit analyses to the most relevant objects.

The 2234 viable sources after the initial inspection stage (see Sect.~\ref{sec:inspection_first_stage}) were distributed amongst team members, such that each source had three independent classifications, and then compiled into a single output catalogue.
For the beam quality flags, we record the consensus flag if all reviewers agreed, or an intermediate classification if any one inspection differed (e.g. if two reviewers considered a spectrum ‘good’, and one ‘poor’, this was recorded as ‘poor/good’).
For the redshifts, we required a single consensus redshift and quality flag.
If reviewers disagreed on the most probable redshift, the object was refitted over a redshift range encompassing all suggested solutions, and reinspected.
Once an agreement had been reached, we refitted 46 objects where the visually estimated redshift differed from \zgrizli, the best-fit redshift from the \grizli\ extraction.
If \zgrizli\ was at the edge of the redshift range used for fitting, we refitted the redshift using a 10\% wider window, to ensure that our results were not biased by the redshift priors.

\section{Results} \label{sec:results}

\subsection{Redshift validation} \label{sec:results_redshift_validation}

\begin{figure*}
    \centering
    \includegraphics[width=\textwidth]{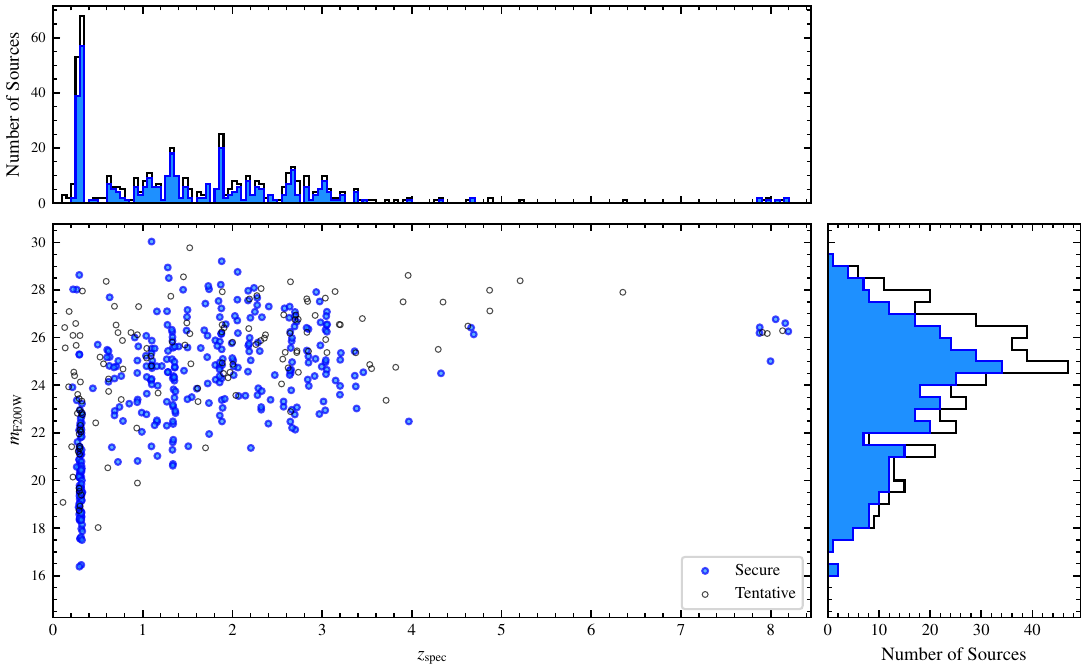}
    \caption{Centre: AB magnitude distribution of all sources (measured in the F200W filter), against the spectroscopic redshift measured from NIRISS. 
    We distinguish between ‘secure’ redshifts, shown as solid blue circles, and ‘tentative’ redshifts, plotted as empty black circles.
    Upper: Corresponding stacked redshift histogram, showing the number of sources in each bin of width 0.05. Here, tentative redshifts are displayed as an unfilled histogram.
    Right: As before, showing the distribution of source magnitudes in bins of width 0.5\,dex.
    }
    \label{fig:mag_z_scatter}
\end{figure*}

\begin{figure} 
    \centering
    \includegraphics[width=\columnwidth]{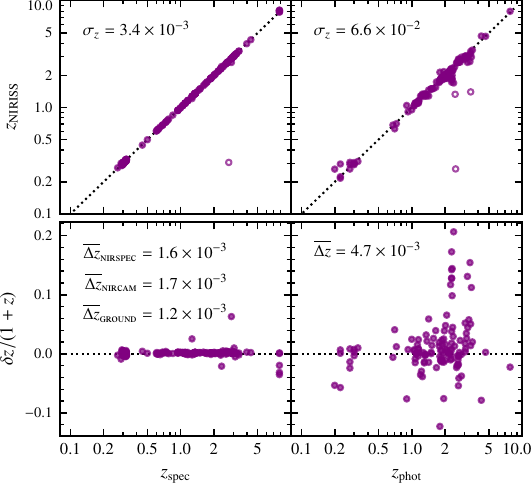}
    \caption{Top: Distribution of NIRISS-derived secure redshifts against the spectroscopic and photometric priors.
    We do not include tentative redshifts here, as these depend on the existing \zphot.
    Bottom: Scaled redshift offset. Here, we adopt the shorthand, $\Delta z=(\zorig-\zniriss)/(1+\zorig)$.
    In the inset text, we calculate the mean offset, separated by the source of \zorig.
    }
    \label{fig:z_niriss_vs_z_spec_phot}
\end{figure}

In Table~\ref{tab:redshift_numbers}, we present the number of secure and tentative redshifts acquired in this analysis, referenced against the existing redshift information for each object (see Sect.~\ref{sec:redshift_priors}).
In total, we derive 488 redshifts, of which 136 are secure redshifts for which either no information or only a photometric redshift was previously available.
In Table~\ref{tab:beam_quality_numbers}, we present the number of 2D spectra that received each quality flag, across both position angles and for all sources in the catalogue.
The quality of the spectra in the F200W blocking filter is substantially lower than in the F115W and F150W filters.
We attribute this to the longer length of the dispersed spectra on the detector, leading to a higher degree of contamination from overlapping sources.

\begin{table} 
\caption{Number of secure and tentative redshifts from this analysis, compared with the existing redshift information from the literature.}
\label{tab:redshift_numbers}
\begin{tabular}{cccc|c}
\hline\hline
          & Had \zspec\ & Had \zphot\ & No $\zphot$ & Total \\
\hline
Secure    &  218  &  129  &   7  &  354  \\
Tentative &   25  &   97  &  12  &  134  \\
\hline
Total     &  243  &  226  &  19  &  488  \\
\hline
\end{tabular}
\end{table}

\begin{table} 
\caption{Number of 2D spectra that received each beam quality classification, including the two position angles.}
\label{tab:beam_quality_numbers}
\begin{tabular}{c|ccc|c}
\hline\hline
Quality Flag & F115W & F150W & F200W & Total \\
\hline
Good & 3210 & 2949 & 2169 & 8328 \\
Good/Poor & 347 & 386 & 481 & 1214 \\
Poor & 352 & 447 & 696 & 1495 \\
Poor/Unusable & 26 & 37 & 61 & 124 \\
Unusable & 3369 & 3485 & 3897 & 10751 \\
\hline
\end{tabular}
\end{table}

We show in Fig.~\ref{fig:mag_z_scatter} the distribution of all these sources in the observed magnitude-redshift plane.
Magnitudes are measured as the sum of the flux within the segmentation map for each source, using the F200W filter, and given in the AB magnitude system. 
Cluster members occupy a narrow region of the magnitude-redshift plane, and are generally the brightest (observed) galaxies in our sample.
As expected, we observe a systematic trend in the apparent magnitude with redshift, with higher redshift galaxies being on average fainter. 
The magnitude distribution peaks at $m_{\rm{AB,F200W}}\approx25$, before dropping off sharply, and we do not measure the redshift of any galaxy apparently fainter than $m_{\rm{AB,F200W}}\geq30$.
In some redshift slices (e.g. $z\approx1.6$), there are no galaxies in our final catalogue, or only tentative redshifts.
Due to the non-overlapping NIRISS filter coverage, the redshift range in which multiple strong lines can be observed is not continuous, as shown by the yellow bars in Fig.~\ref{fig:wavelength_coverage}.
The wavelength coverage also drives the sudden drop in sources at $z\gtrsim3.4$, until $z\approx7.3$, when \ion{Ly}{$\alpha$} enters the F115W filter.

In Fig.~\ref{fig:z_niriss_vs_z_spec_phot}, we compare our secure redshifts against the previous spectroscopic and photometric redshifts, \zorig, where the dotted lines represent a perfect agreement (i.e. $\zorig=\zniriss$).
We measure an overall scatter of just $\sigma_{\zspec}=3.4\e{-3}$ with regard to previous spectroscopic redshifts, and $\sigma_{\zphot}=6.6\e{-2}$ for photometric redshifts.
We identify four sources for which $\lvert\zniriss-\zorig\rvert>1$, shown as open circles in Fig.~\ref{fig:z_niriss_vs_z_spec_phot}.
\begin{itemize}
    \item ID 3514 is the only discrepant source for which the existing redshift was derived from spectroscopy.
    It was observed with NIRSpec as part of the GLASS-JWST ERS programme (1324), with MSA\_ID 410074 \citep{mascia_glass-jwst_2024}, and an estimated redshift $z=2.736$.
    In our NIRISS observations, we identify the presence of the molecular absorption bands (see Sect.~\ref{sec:cluster_galaxies}), and a tentative detection of \Siii\ emission, giving $\zniriss=0.305$.
    Refitting the NIRSpec spectrum at this redshift, we also identify emission at wavelengths corresponding to the \ion{Pa}{$\alpha$} and \ion{Br}{$\beta$} lines.
    As the observed wavelengths of this combination of emission lines cannot be reproduced by any higher-redshift source, we are confident that the quoted NIRSpec redshift is due to a misidentification of a star-forming region in a galaxy at the cluster redshift.
    We therefore consider $\zniriss=0.305$ for this source to be a secure redshift.
    \item ID 321 was estimated to be a $\zphot=2.48$ source in the ASTRODEEP photometric catalogue. 
    In the NIRISS spectrum, we are able to clearly identify the \ion{H}{$\beta$}, \Oiii, and \ion{H}{$\alpha$} lines, securing a spectroscopic redshift at $z=1.33$.
    The direct imaging for this object reveals an extremely luminous point source, at the centre of a faint disc.
    We therefore suggest that this galaxy is the origin of the X-ray point source visible in previous studies of the cluster \citep{kempner_chandra_2004,merten_creation_2011,owers_dissection_2011}, and contributions from this probable active galactic nucleus (AGN) may have biased the photometric redshift. 
    We will present a more detailed look at this source in a future analysis.
    \item ID 1990 is a very compact galaxy, previously assigned $\zphot=0.025$. 
    We find clear evidence of spatially offset star formation, securing $\zspec=2.05$ via the \Halpha\ and \Oiii\ lines.
    \item ID 3499 is another very compact source, with a previous $\zphot=2.5$.
    We instead identify \PaB, \ion{Pa}{$\gamma$}, \ion{He}{I}-1083, and \Siii, marking this as a foreground interloper at $z=0.265$.
    The source itself appears towards the edge of a disc galaxy (ID 3500) along the line of sight, although the grism coverage prevents us from deriving a secure redshift for this disc galaxy.
    We therefore cannot determine if these objects are physically associated based on the available data, and we do not speculate further about their nature in this study.
\end{itemize}

In the bottom row of Fig.~\ref{fig:z_niriss_vs_z_spec_phot}, we show the redshift deviation in the form $\Delta z = \delta z/(1+z)$, where $\delta z = \zorig-\zniriss$.
For sources with previous spectroscopic measurements, we find a median deviation of $\Delta z = 1.3\e{-3}$, with scatter $1.6\e{-3}$.
Whilst this offset in $\Delta z$ is less than the scatter, it is persistent, even when separating galaxies by the origin of the previous redshift (see the inset text in Fig.~\ref{fig:z_niriss_vs_z_spec_phot}).
We attribute this to deficiencies in the wavelength calibration of NIRISS.
Whilst substantial improvements have been made over the pre-launch calibrations \citep{matharu_updated_2022,pirzkal_next_2024}, small systematic offsets (from external measurements, and between grism filters) have also been recorded in other fields using the \texttt{*221215.conf} trace configuration files (PASSAGE team, private communication; \citealt{rihtarsic_canucs_2025}).
We therefore suggest that the redshifts derived in this work should be considered accurate only to the level of $\Delta z\approx1\e{-3}$.
We note that there are a small number of sources displayed in Fig.~\ref{fig:z_niriss_vs_z_spec_phot} with offsets of $\Delta z\geq0.02$.
With the exception of those at $z\geq7$, which are discussed further in Sect.~\ref{sec:overdensities}, we are confident in the accuracy of the NIRISS-derived redshifts presented here.
In particular, two of these galaxies (IDs 1511 and 1933) were independently reduced by \cite{marchesini_early_2023} using an earlier version of the data, and our results are consistent with theirs.

Comparing against the photometric redshifts from the ASTRODEEP catalogue, we find our results are in relatively good agreement.
The median deviation is $\Delta z=4.7\e{-3}$, consistent with the systematic positive offset seen for spectroscopic observations, although with a much greater scatter of $\sigma_{\Delta z}=2.5\e{-2}$.
From the distribution in Fig.~\ref{fig:z_niriss_vs_z_spec_phot}, we see possible signs of discretisation noise in the previous $\zphot$, with five galaxies at $\zphot\approx2.3$ having a deviation $\Delta z\geq0.1$, although this is a relatively minor effect.
As a direct comparison to \cite{merlin_astrodeep-jwst_2024}, we compute the fraction of outliers $f_{\rm{outliers}}$, the fraction of sources with $\Delta z>0.15$, for all of the 226 secure and tentative redshifts derived in this work. 
We find $f_{\rm{outliers}}=5.4\%$, confirming the very high accuracy of the original catalogue.

\subsection{Catalogue presentation}

We describe here the content of the redshift catalogue.
The released catalogue includes all 3652 sources present in the final extraction, in order to distinguish between sources that were not extracted, and those for which we could not determine the redshift.

\begin{itemize}
    \item ID\_NIRISS: The assigned ID in our data reduction, drawn from the composite segmentation map described in Sect.~\ref{sec:source_detection}. This ID is unique to this analysis.
    \item RA; DEC: The world coordinates of the flux-weighted centre of the object, given in the ICRS.
    \item Z\_NIRISS: The best-fit redshift to the NIRISS 2D and 1D spectra, as determined by multiple team members following a redshift fit by \grizli.
    \item Z\_FLAG: The redshift quality flag, described in more detail in Sect.~\ref{sec:inspection_second_stage}. This is given as an integer, with the following meanings:
    \begin{itemize}
        \item 0 : The object was not extracted.
        \item 1 : The object was rejected in the initial inspection (see Sect.~\ref{sec:inspection_first_stage}).
        \item 2 : The object redshift could not be determined in the second stage of the visual inspection.
        \item 3 : The redshift was flagged as ‘tentative’.
        \item 4 : The redshift was determined to be ‘secure’.
    \end{itemize}
    \item \{filter\}\_\{angle\}\_FLAG: The quality flags for each combination of filter (e.g. ‘F150W’) and position angle (e.g. ‘341.0’), as determined by visual inspection of multiple team members. These have the following meanings:
    \begin{itemize}
        \item 0 : The beam was considered to be ‘Unusable’.
        \item 1 : The beam quality was ‘Poor’.
        \item 2 : The beam was ‘Good’ (the default classification).
    \end{itemize}
    Intermediate classifications were allowed where there was no clear consensus, (e.g. $\rm{F115W\_341.0\_FLAG}=1.5$ would indicate that the beam was rated as both ‘Good’ and ‘Poor’ in different inspections).
    \item N\_LINES: The number of emission lines detected by \grizli\ when fitting the 2D spectra.
    \item NAME\_LINES: The names of the emission lines, given as a semicolon-separated string, (e.g. ‘Ha;OIII;Hb;OII’).
    \item FLUX\_\{line\}; ERR\_\{line\}: The line flux and associated error as measured by \grizli\ on the 2D spectra. The line name is the same as given in \textsc{name\_lines}.
\end{itemize}

We wish to highlight that due to the spectral resolving power of NIRISS, it is not possible to separate the flux for particular combinations of lines. Additionally, the default \grizli\ reduction combines the flux for certain elements. As a result,
\begin{itemize}
    \item FLUX\_OII is the combined flux from the \Oiilam\ doublet.
    \item FLUX\_OIII is the combined flux from both the $\Oiii{\lambda 4959}$ and $\Oiii{\lambda 5007}$ emission lines.
    \item FLUX\_Ha represents the combined flux from the \Halpha, $\Nii{\lambda 6548}$, and $\Nii{\lambda 6583}$ complex.
    \item FLUX\_SII measures the flux from both the $\Sii{\lambda 6716}$ and $\Sii{\lambda 6731}$ lines.
\end{itemize}

\subsection{First insights from the catalogue}

\subsubsection{Cluster galaxies} \label{sec:cluster_galaxies}

\begin{figure}  
    \centering
    \includegraphics[width=\columnwidth]{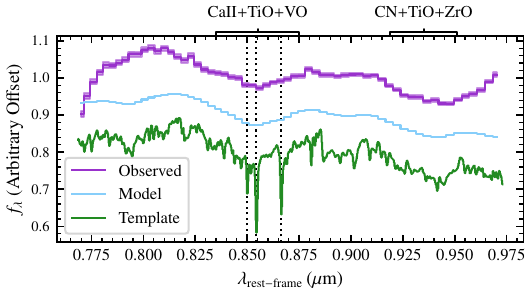}
    \caption{Example of the rest-frame absorption features visible in a number of the passive cluster members. At the top, we display the observed 1D spectrum for galaxy ID 3528 in the F115W filter, after de-redshifting the spectrum. At the bottom we show the best-fit template spectrum at the native template resolution, and in the middle the best-fit template after forward-modelling the NIRISS spectral resolution and the morphological broadening effect of this particular galaxy. The dotted lines indicate the \ion{Ca}{II} triplet, and we annotate at the top the approximate area covered by the molecular absorption bands. The fall-off and upturn at the blue and red outer edges of the observed 1D spectrum are merely an artefact of the \cite{horne_optimal_1986} 1D extraction algorithm for extended sources (see also Fig.~\ref{fig:redshift_quality_examples}c). This does not affect the redshift fit, which is performed on the original 2D spectra.}
    \label{fig:absorption_bands}
    
\end{figure}

As seen in Fig.~\ref{fig:mag_z_scatter}, the redshift distribution of our catalogue shows a very strong peak centred on the cluster redshift ($z=0.3064$). We measure the redshift of 113 galaxies within the range $0.28\leq z<0.33$, of which 17 have no previous spectroscopic coverage. However, from Fig.~\ref{fig:z_niriss_vs_z_spec_phot}, we also observe that this redshift range has an increased scatter in the deviation, compared to higher-redshift galaxies ($\sigma_{\Delta z}=3.4\e{-3}$ \textit{vs.} $\sigma_{\Delta z}=9.8\e{-4}$). We attribute this to several distinct factors. Firstly, there are no strong emission lines visible in the grism filters at this redshift (see Fig.~\ref{fig:wavelength_coverage}), which restricts our ability to precisely determine the redshift even for active star-forming galaxies. Furthermore, the vast majority of cluster members in this field are passive, with redshifts determined almost exclusively by absorption features in the 0.8-0.9$\,\mu$m rest-frame range, which are much broader and thus harder to fit than narrow emission lines.

Secondly, the cluster members are on average larger than higher redshift galaxies. This leads to a severe ‘morphological broadening’ effect \citep{he_early_2024}, further reducing the effective resolution of the 1D equivalent spectra. Whilst this does not affect the \grizli\ fit, which forward models the full 2D spectra, it limits our ability to verify the redshift to a high degree of precision during the visual classification stage (see Sect.~\ref{sec:inspection_second_stage}). We show an example spectrum in Fig.~\ref{fig:absorption_bands}, against the best-fit high-resolution template spectrum before and after forward-modelling to match the observations. The effective resolving power of $R\ll150$ at these wavelengths washes out all narrow features, leaving only a broad, but characteristic ‘double-dip’ pattern between $1.0\lesssim\lambda_{\rm{obs}}\lesssim1.3$. Lastly, there is a distinct possibility that the template set used in \grizli\ for redshift fitting does not contain enough old or high-metallicity models to allow for an accurate fit in this region of the spectrum. This specific issue has been highlighted before by \cite{noirot_first_2023}, although such deficiencies are not readily apparent from a visual inspection of the fitted spectra.

\begin{figure*} 
    \centering
    \includegraphics[width=\textwidth]{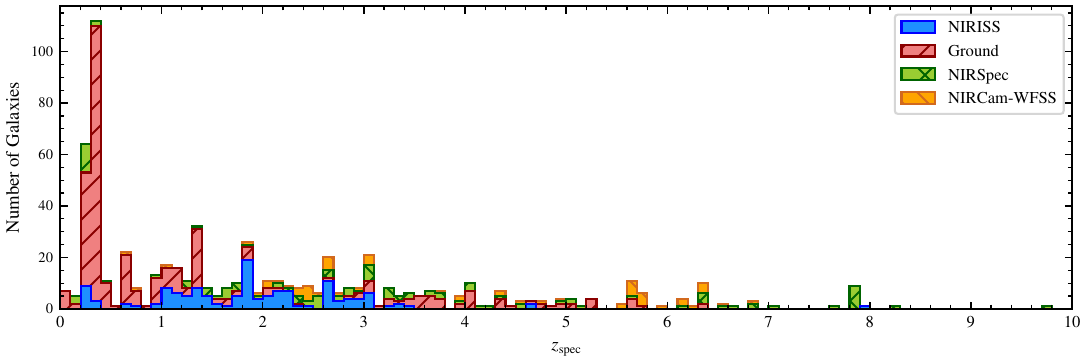}
    \caption{Histogram of all spectroscopically confirmed redshifts, within the NIRISS field of view. Here, we plot only the ‘secure’ NIRISS redshifts, and redshifts from ALT where at least two lines were detected. Many galaxies are observed by multiple instruments, although they are only included once in this figure and assigned in the order: ‘ground’, NIRSpec, NIRCam-WFSS, and NIRISS. The blue region on this histogram thus corresponds to the distribution of new secure redshifts presented in this catalogue.}
    \label{fig:full_z_hist}
\end{figure*}

\subsubsection{Background overdensities} \label{sec:overdensities}

In Fig.~\ref{fig:full_z_hist}, we present the full spectroscopic redshift distribution within the NIRISS field of view, cross-matching all prior catalogues described in Sect.~\ref{sec:redshift_priors}. We assign a source based on the instrument with which each object was first observed (in the order ‘Ground’, NIRSpec, NIRCam-WFSS, and NIRISS, where we grouped all ground-based instruments together). We show only ‘secure’ redshifts, as defined by the relevant quality flags in each constituent catalogue, and as such, the NIRISS portion of Fig.~\ref{fig:full_z_hist} highlights the distribution of new secure redshifts derived in this work.

\begin{figure}
    \centering
    \includegraphics[width=\columnwidth]{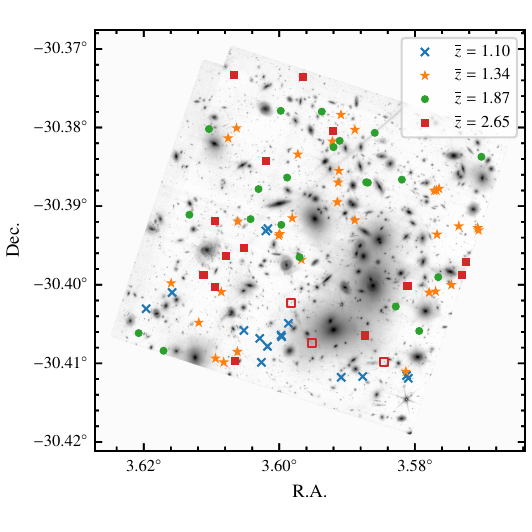}
    \caption{On-sky distribution of sources in the four overdensities at $\overline{z}=1.10$, 1.34, 1.87, and 2.65, marked by blue crosses, orange stars, green circles, and red squares, respectively. Open red squares denote the multiple image system discussed in Sect.~\ref{sec:multiple_images}. In the background, we show the NIRISS direct imaging in the F200W filter.}
    \label{fig:overdensities_sky_plot}
\end{figure}

\begin{figure}
    \centering
    \includegraphics[width=\columnwidth]{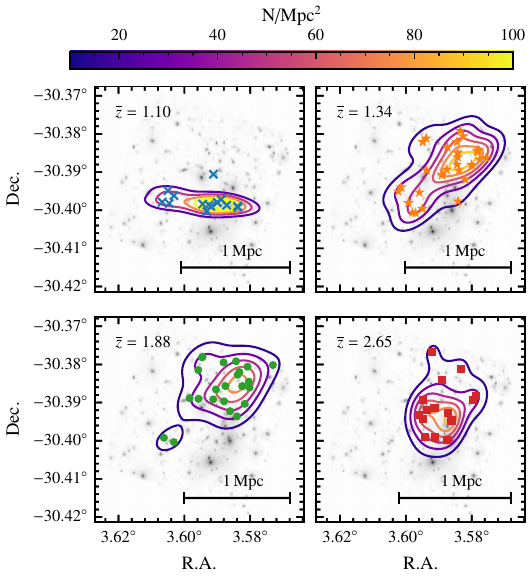}
    \caption{
  Source plane distribution of galaxies in the four overdensities, reconstructed using the cluster lensing model of \cite{bergamini_glass-jwst_2023}.
    The markers for each overdensity are the same as in Fig.~\ref{fig:overdensities_sky_plot}.
    We overlay contours of the number surface density at each redshift, derived using Gaussian kernel-density estimation.
    }
    \label{fig:od_source_plane}
\end{figure}

Whilst we can identify multiple peaks and, hence, potential overdensities in the overall redshift distribution behind the cluster, we focus here on those that are also visible in the NIRISS data only (Fig.~\ref{fig:mag_z_scatter}). We locate four such overdensities, each with $\geq10$ sources with secure spectroscopic redshifts, at $z\approx1.10$, 1.34, 1.87, and 2.65, with the observed locations of the galaxies shown in Fig.~\ref{fig:overdensities_sky_plot}. The overdensities at $\overline{z}\approx1.10$ and $\overline{z}\approx2.65$ were not previously identified in the literature, to the best of our knowledge, and the significance of the peak at $\overline{z}\approx1.87$ has increased substantially as a result of the NIRISS data. In Fig.~\ref{fig:od_source_plane}, we reconstruct the positions of the galaxies in the source plane for each overdensity, using the cluster lens model of \cite{bergamini_glass-jwst_2023}.

We find 15 sources in the range $1.09\leq z <1.11$ ($\overline{z}=1.10$), of which 13 are included in our NIRISS catalogue. Two of these sources (IDs 521 and 561) potentially form a single interacting system, or may instead represent a star-forming clump that has been de-blended from the host galaxy during the source detection stage of our analysis (see Sect.~\ref{sec:source_detection}). We do not identify any multiply imaged systems within this redshift range. The spatial distribution of these galaxies is shown in Fig.~\ref{fig:overdensities_sky_plot} by the blue crosses. With the exception of a pair of galaxies near the centre of the NIRISS field of view, all other sources are close to the southern edge of the cluster. However, the reconstructed source plane distribution in Fig.~\ref{fig:od_source_plane} shows that the majority of galaxies are grouped behind the cluster core. With an estimated number surface density exceeding $10^2/\rm{Mpc}^2$, we suggest that this overdensity may be associated with a singular group, although we defer a full discussion of this to future analyses.

We identify 28 sources in the range $1.32\leq z <1.37$ ($\overline{z}=1.34$) using NIRISS, and an additional 4 sources from the ancillary redshift catalogues. These include an interacting pair (IDs 1689 and 1718, \citealt{wang_census_2020}), a probable AGN (ID 321, Sect.~\ref{sec:results_redshift_validation}), and multiple galaxies with resolved spiral structure (IDs 1853, 3070, and 3234). It does not appear that any of these galaxies are multiple images. The galaxies within this redshift range appear to be uniformly distributed over the NIRISS field of view, perhaps indicating a chance alignment of a large-scale overdensity with the cluster core. Within the source plane, we observe a clustering of galaxies aligned to the north-west of the cluster core, with a similar density to those at $\overline{z}=1.10$. There also appears to be a secondary grouping of galaxies closer to the centre of the field.

We find 20 sources in the range $1.85\leq z <1.90$ ($\overline{z}=1.87$) using NIRISS, out of a total of 21 sources including the ancillary data. Two sources (IDs 2355 and 2363) are very closely associated and may be star-forming clumps in the same galaxy, or a post-merger remnant. While the majority of sources at this redshift are compact, we observe three lensed arcs (IDs 867, 1221, and 2408), and one with multiple resolved star-forming clumps (ID 2938). In Figure~\ref{fig:overdensities_sky_plot}, we see that there is a slight preference for galaxies in this redshift range to be located on the northern side of the cluster core. Reconstructing the positions of these galaxies within the source plane, we observe a fairly scattered group covering $\approx1\,\rm{Mpc}^2$.

We identify 18 sources in the range $2.63\leq z <2.67$ ($\overline{z}=2.65$), 15 of which were extracted with NIRISS. Of these, three correspond to a new multiple image system discussed in Sect.~\ref{sec:multiple_images}. Although this artificially inflates the number of galaxies, we note that 16 unique sources is still a significant overdensity in this redshift slice. One object (ID 1511) is a previously identified quiescent galaxy \citep{marchesini_early_2023}, displaying a strong Balmer break and NaD absorption, whilst ID 2663 shows extended spiral structure. ID 3051 was previously identified as a dusty star-forming galaxy, based on ALMA 1.1\,mm continuum detections \citep[A2744-ID07, ][]{gonzalez-lopez_alma_2017}, and we detect here the corresponding \Oiii\ emission. These galaxies do not appear to be preferentially located in any part of the observed field of view, but the source plane reconstruction reveals a tighter grouping aligned with the cluster core. Given the large deflection angles at this redshift, it is probable that there are additional sources located outside of the NIRISS field of view, which will be investigated further in a future analysis.

As indicated in Fig.~\ref{fig:wavelength_coverage}, all four of these overdensities are aligned with redshift ranges where two or more strong emission lines (\ion{H}{$\alpha$}, \Oiii, and \Oii) have simultaneous grism filter coverage.
We consider the possibility that instrumental effects have biased the redshift distribution (noting also that \citealt{noirot_first_2023} reported overdensities at $\overline{z}=1.08$, 1.37, and 1.98 with NIRISS observations of a similar cluster).
Whilst these overdensities are in advantageous redshift ranges for detection by NIRISS, the sources themselves occupy a much narrower redshift range than would be expected if driven solely by the filter coverage. 
Indeed, at $z\approx1.34$ and 1.87, these distributions are clearly visible as vertical stripes in the redshift-magnitude plane of Fig.~\ref{fig:mag_z_scatter}.
In addition, of the four overdensities discussed here, at least two have been identified in previous studies -- the peak at $\overline{z}=1.35$ is visible in \cite{wang_grism_2015}, and the peak at $\overline{z}=1.87$ in \cite{naidu_all_2024}, albeit driven by tentative single-line detections (and hence not shown in Fig.~\ref{fig:full_z_hist}).
We are therefore confident that whilst the instrument design may influence our ability to detect these overdensities, we are nevertheless observing real peaks in the redshift distribution.

One final point here relates to galaxies at $\zniriss\geq7.8$.
There is a known overdensity at $z\approx7.9$ in the A2744 field, corresponding to the second-most distant protocluster known to date, and discussed in detail in various other studies \citep{zheng_young_2014,roberts-borsani_early_2022,morishita_early_2023,morishita_metallicity_2025}.
We observe a number of these galaxies with NIRISS.
Those with secure redshifts and existing \zspec\ include ID 2514 (YD8), ID 3009 (ZD3), ID 3013 (ZD6), ID 3062 (ZD2/GLASSz8-1), and ID 3162 (GLASSz8-2).
There is a considerable scatter between these previous redshifts (derived using NIRSpec) and those in our catalogue, seen in Fig.~\ref{fig:z_niriss_vs_z_spec_phot}.
We attribute this to the low continuum S/N obtained with NIRISS, which combined with the relatively low spectral resolution, makes a precise location of the Lyman break difficult in our observations.
To these previously known high-redshift galaxies, we add three with tentative redshifts (IDs 1839, 3594, and 3703), all located $\geq\ang{;;90}$ away from the protocluster core in the image plane.
We also observe a new galaxy (ID 2992), with a secure $\zniriss=8.0$.
This is located $\approx\ang{;;27}$ from the protocluster core in the image plane, and will be discussed further in a future analysis.

\subsubsection{Multiple images} \label{sec:multiple_images}

\begin{figure}  
    \centering
    \includegraphics[width=\columnwidth]{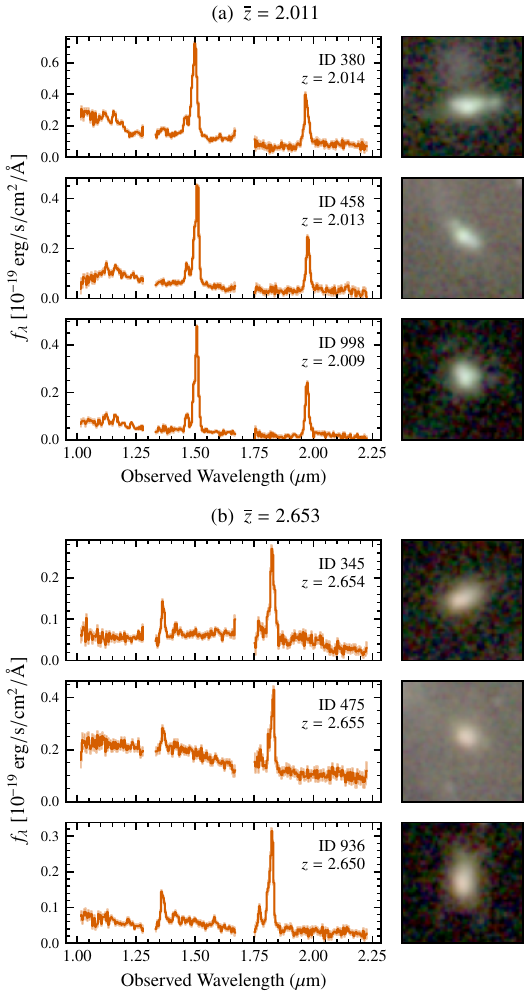}
    \caption{Direct image cut-outs and spectra for (a) the previously observed multiple image and (b) the photometric candidate multiple image, for which we obtain a spectroscopic redshift. Left column: Extracted 1D spectra for each source, across all available filters. In the inset, we show the ID and best-fit \grizli\ redshift. Right column: Direct image cut-outs for each source, measuring \ang[angle-symbol-over-decimal]{;;1.5} per side. We create RGB images by mapping the NIRISS F115W, F150W, and F200W imaging to the blue, green, and red channels, respectively.}
    \label{fig:multiple_images_cutouts}
\end{figure}

A large sample of multiply imaged sources with reliable redshifts is vital for the construction of high-precision cluster mass models. Abell 2744 is an extremely well-studied strong lensing system \citep[e.g.][]{mahler_strong-lensing_2018,richard_atlas_2021,bergamini_glass-jwst_2023}, and the 488 sources in our redshift catalogue do not necessarily correspond to unique galaxies. To identify multiple images, we rely on the catalogue compiled by \cite{cha_precision_2024}. Their analysis separated multiply imaged systems into two categories, gold and silver class. Gold-class images were defined as those with secure spectroscopic redshifts, and silver-class images as those with consistent photometric redshifts across multiple papers \citep[e.g.][]{jauzac_extraordinary_2016,mahler_strong-lensing_2018,bergamini_new_2023,bergamini_glass-jwst_2023,furtak_uncovering_2023}. After cross-matching against our NIRISS catalogue, we identify 12 sources in common:
\begin{itemize}
    \item Three sources (IDs 345, 475, and 936) were previously recognised as a silver-class multiply imaged galaxy. 
    We confirm the spectroscopic redshift of this system for the first time, at $\overline{z}=2.653\pm0.002$, with the redshifts of each source measured as $\zniriss=2.654$, 2.655, and 2.650, respectively.
    \item An additional three sources (IDs 380, 458, and 998) in our catalogue correspond to a single gold-class multiple-image system. 
    The previous $\zspec=2.017$ is in good agreement with our own measurement of $\zniriss=2.012\pm0.002$ (with individual redshifts of $\zniriss=2.014$, 2.013, and 2.009), considering the systematic offset discussed in Sect.~\ref{sec:results_redshift_validation}.
    \item Six sources in our catalogue (IDs 749, 867, 3591, 3615, 3884, and 3890) are each associated with separate multiply imaged systems, with gold classifications (each system had a previous spectroscopic redshift).
    However, although all of these sources have one or more counter-images within the NIRISS field of view, we only obtain reliable redshifts for these specific IDs from our observations.
\end{itemize}

We present in Fig.~\ref{fig:multiple_images_cutouts} cut-outs of these confirmed multiple image systems present in our redshift catalogue, alongside the extracted 1D spectrum and best-fit redshift for each source. We note that identification of multiple images from NIRISS data without prior knowledge of their location remains extremely difficult. Small imperfections in the contamination model can affect the shape of the extracted continuum, and counter-images close to the cluster core are likely to be washed out by the dispersed ICL. We therefore prefer not to speculate on any new multiple images in this catalogue.

\section{Conclusions} \label{sec:conclusions}

This study presents a comprehensive redshift catalogue, derived using the NIRISS observations of the GLASS-JWST ERS programme in the Abell 2744 cluster field.
We summarise our work as follows:

\begin{enumerate}
    \item We determine 488 spectroscopic redshifts, spanning a range $0.1\leq z \leq 8.2$.
    Those flagged as ‘secure’ show excellent agreement with previous measurements, with a scatter of  $\sigma_z=3.4\e{-3}$ compared to spectroscopic redshifts and $\sigma_z=6.6\e{-2}$ compared to photometric redshifts. 
    We observe a small systematic offset in both cases, with a median $\Delta \zspec = (1.3\pm1.6)\e{-3}$ and $\Delta \zphot = (4.7\pm25)\e{-3}$.
    We interpret this as due to small residual errors in the NIRISS wavelength calibration.
    \item We derive secure (tentative) redshifts for 136 (109) galaxies, which had either no previous existing redshift or only a photometric redshift.
    These include 17 galaxies at the cluster redshift, $0.28\leq z \leq0.33$, and one galaxy at $z\approx8$.
    \item We associate three sources (IDs 345, 475, and 936) with a previous photometrically identified multiply imaged galaxy. 
    We confirm the spectroscopic redshift of this system for the first time as $z=2.653\pm0.002$.
    \item We combine our new redshift catalogue with existing catalogues from the literature, in order to show the full spectroscopic redshift distribution within the NIRISS field-of-view.
    We identify four distinct overdensities, visible also in the NIRISS data alone, at $\overline{z}=1.10$, 1.34, 1.87, and 2.65.
    Galaxies around $\overline{z}=1.10$ appear almost exclusively in the south of the cluster, whereas those at higher redshifts show a more uniform distribution across the field.
    \item We present a visualisation and classification tool \href{https://pypi.org/project/pyGCG/}{\pygcg}, aimed at simplifying the inspection of both 2D and 1D spectra reduced using \grizli and tested against data products from both this analysis and other NIRISS programmes such as PASSAGE \citep{malkan_passage-parallel_2021}.
    \item We also present a forced extraction tool, \href{https://pygrife.readthedocs.io/}{\pygrife}, which allows for the extraction of arbitrary on-sky regions in reduced grism data.
\end{enumerate}

This catalogue will enable improvements to existing cluster mass models, and pave the way for future analyses focussing on the spatially resolved properties of galaxies around cosmic noon.

\section*{Data availability}

The catalogue is available at the CDS via anonymous ftp to \href{https://cdsarc.cds.unistra.fr}{cdsarc.cds.unistra.fr} (\href{ftp://130.79.128.5/}{130.79.128.5}) or via \href{https://cdsarc.cds.unistra.fr/viz-bin/cat/J/A+A/}{https://cdsarc.cds.unistra.fr/viz-bin/cat/J/A+A/}.
The data used are available at the MAST, with JWST programme ID 1324.
The catalogue will also be available at MAST, as part of the GLASS-JWST High-Level Science Product collection\footnote{\url{https://archive.stsci.edu/hlsp/glass-jwst}}.
The source code for the classification GUI, \pygcg, is available on GitHub\footnote{\url{https://github.com/PJ-Watson/pyGCG}} and can be installed directly from the Python Package Index\footnote{\url{https://pypi.org/project/pyGCG/}}.
The source code for the forced extraction tool, \pygrife, is available on GitHub\footnote{\url{https://github.com/PJ-Watson/PyGriFE}}.

\begin{acknowledgements}
    We thank the anonymous referee whose useful comments helped us to improve the manuscript. The data were obtained from the MAST at the Space Telescope Science Institute, which is operated by the Association of Universities for Research in Astronomy, Inc., under NASA contract NAS 5-03127 for JWST. 
    These observations are associated with programme JWST-ERS-1324. 
    We acknowledge financial support from NASA through grants JWST-ERS-1324.
    We also acknowledge support from the INAF Large Grant 2022 `Extragalactic Surveys with JWST' (PI Pentericci). 
    P. W., B.V. and A.A. are supported by the European Union – NextGenerationEU RFF M4C2 1.1 PRIN 2022 project 2022ZSL4BL INSIGHT. 
    P.W. and B.V. acknowledge support from the INAF Mini Grant `1.05.24.07.01 RSN1: Spatially Resolved Near-IR Emission of Intermediate-Redshift Jellyfish Galaxies' (PI Watson).
    MB acknowledges support from the ERC Grant FIRSTLIGHT and from the Slovenian national research agency ARIS through grants N1-0238 and P1-0188.
    X.W. is supported by the National Natural Science Foundation of China (grant 12373009), the CAS Project for Young Scientists in Basic Research Grant No. YSBR-062, the Fundamental Research Funds for the Central Universities, the Xiaomi Young Talents Program, and the science research grant from the China Manned Space Project.
    D.M. and C.T. gratefully acknowledge funding from HST-GO-17231.
    We also thank Jasleen Matharu for helpful discussions on working with \grizli, the PASSAGE team for discussions on wavelength calibrations, and the GLASS team.
\end{acknowledgements}

\bibliographystyle{aa} 
\bibliography{references} 

\begin{appendix}

\section{Grism classification GUI} \label{app:pygcg}

\begin{figure*}
    \includegraphics[width=\textwidth]{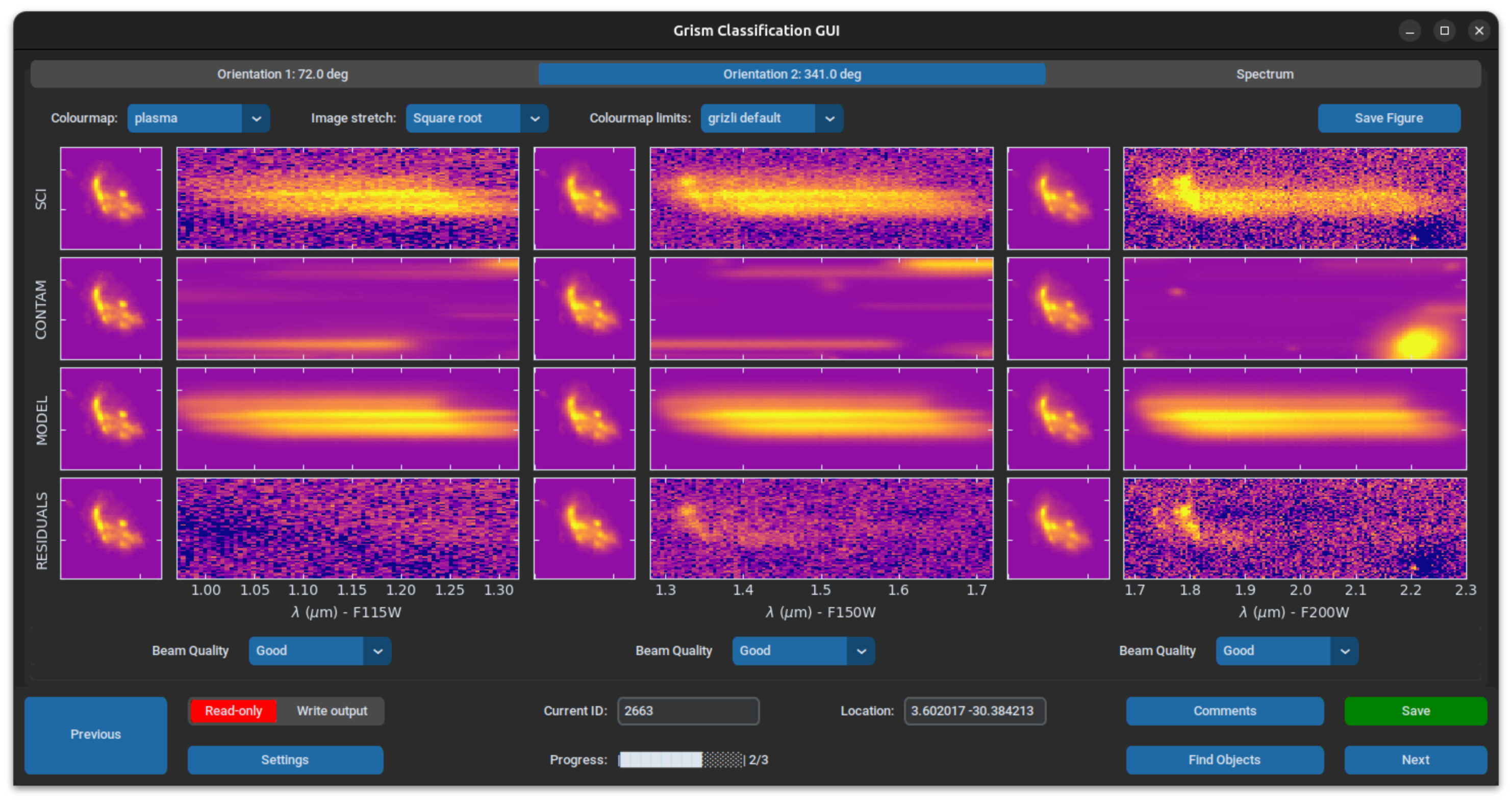}
    
    \includegraphics[width=\textwidth]{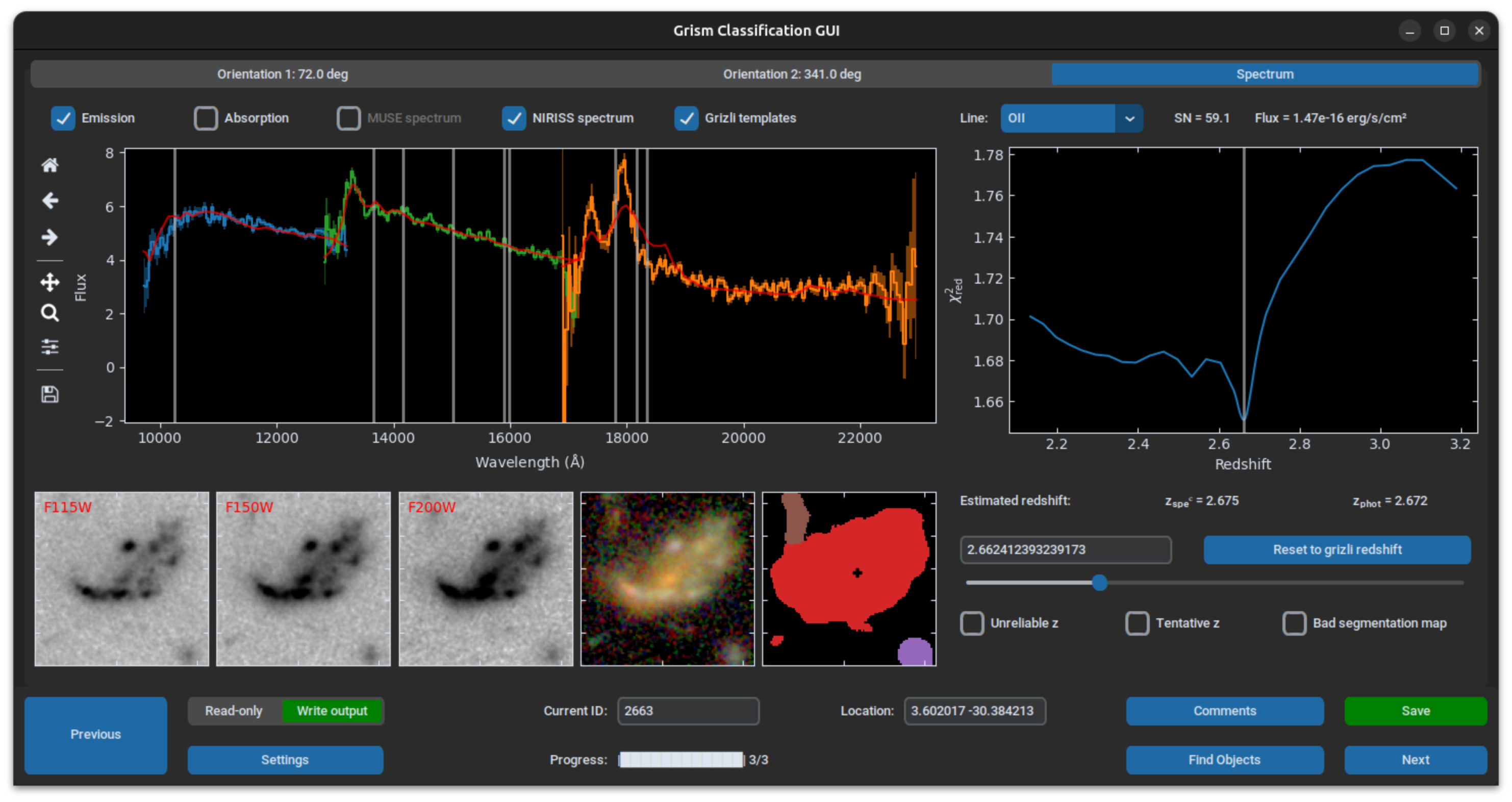}
    \caption{Top: Overview of one of the \pygcg\ tabs, displaying the stacked beams for a single position angle. 
    From top to bottom, we show the dispersed 2D `science' spectrum for the object of interest, the contaminating flux from nearby objects already subtracted from the science spectrum, the continuum model from the best-fitting combination of templates, and the residuals from subtracting the continuum model from the science spectrum.
    This lower row should reveal any emission lines if present, as evident here in the F150W and F200W filters.
    Bottom: Equivalent tab for the extracted 1D spectrum, showing also the direct image cut-outs (if available), a composite RGB image, and the segmentation map (where the object of interest is always coloured in red).
    Whilst more compact galaxies often have consistent 1D and 2D spectra, this is not always the case for extended objects.
    As shown here, spatially offset emission can result in misleading wavelength solutions in the extracted 1D spectrum.
    The plot of $\chi^2$ against redshift on the right-hand panel reveals that this is merely an artefact of the extraction algorithm, and the returned \zgrizli\ is the most probable redshift solution.
    }
    \label{fig:pygcg_screenshots}
\end{figure*}

\pygcg\ is a cross-platform GUI, and is designed as a generalised successor for the GLASS Inspection GUI, \textsc{gig(z)} \citep{treu_grism_2015}, built using the \textsc{tkinter} framework for \textsc{python}.
The source code is available at \href{https://github.com/PJ-Watson/pyGCG}{github.com/PJ-Watson/pyGCG}, and it can also be installed directly via the \href{https://pypi.org/project/pyGCG/}{Python Package Index}.
\pygcg\ combines both the inspection and classification functionality of its predecessor and allows users to interactively browse a selection of the reduced data products from \grizli, with the option of also writing classifications to a table.
At a minimum, \pygcg\ requires the \texttt{*\_stack.fits} and \texttt{*\_full.fits} files, or equivalent, along with an input catalogue.
Additional functionality can be enabled if the segmentation map and direct images are also provided.

In the first two tabs of the GUI, one of which is shown in Fig.~\ref{fig:pygcg_screenshots}a, \pygcg\ displays the dispersed 2D spectra for each position angle (or grism orientation), across all filters.
The display includes the contamination-subtracted science spectrum, the contamination model, the best-fit continuum model for the object of interest, and the residual emission lines.
Whilst these plots are analogous to the \texttt{*\_stack.png} files produced by \grizli, they are fully interactive and allow users to change the colour map, stretch, and colour limits using the drop-down menus provided at the top of the window.
These figures can also be saved as \texttt{*.png} files for later reference.
Below the beam plots, a drop-down menu is provided for each filter, allowing for the classification of all combinations of the filters and grism orientations.

In the third tab of the GUI, shown in Fig.~\ref{fig:pygcg_screenshots}b, the focus is on the optimally extracted 1D spectrum.
Below, if available, cut-outs in each filter are used to create a composite RGB image, alongside the segmentation map for the relevant object.
The right-hand side of the panel displays a plot of $\chi^2$ against \zgrizli.
Beneath this is a redshift slider, allowing for an interactive display of the location of emission lines, overplotted on the 1D spectrum, based on a pre-supplied configuration file.
This enables an efficient search for the best-fit redshift, in case of multiple possible solutions, or if the fitted redshift range does not cover the correct solution.

\section{Forced extraction of pre-processed grism data}\label{app:pygrife}

We developed a forced extraction tool, \pygrife, intended to be used for re-extracting regions from pre-processed grism data \citep{treu_glass-jwst_2022}, and released to the community as a \textsc{python} package, with documentation available at \href{https://pygrife.readthedocs.io/}{pygrife.readthedocs.io/}.
The full reduction pipeline for NIRISS data can be extremely time-consuming and computationally expensive, depending on both the density of the field and the pipeline parameters chosen.
In particular, for sources which already exist in the contamination map but require further de-blending (or recombination), regenerating the full contamination model is a rather inefficient process.
This package hooks into various \grizli\ functions and allows for the extraction of arbitrary regions through a localised change to the existing contamination map. 
This allows the extraction functions to be fully compatible with more complicated contamination models, including those with physically motivated spectra for each individual source.

\section{Examples: Beam quality} \label{app:examples_beam_quality_flag}

\begin{figure*}
    \centering
    \includegraphics[width=0.8\textwidth]{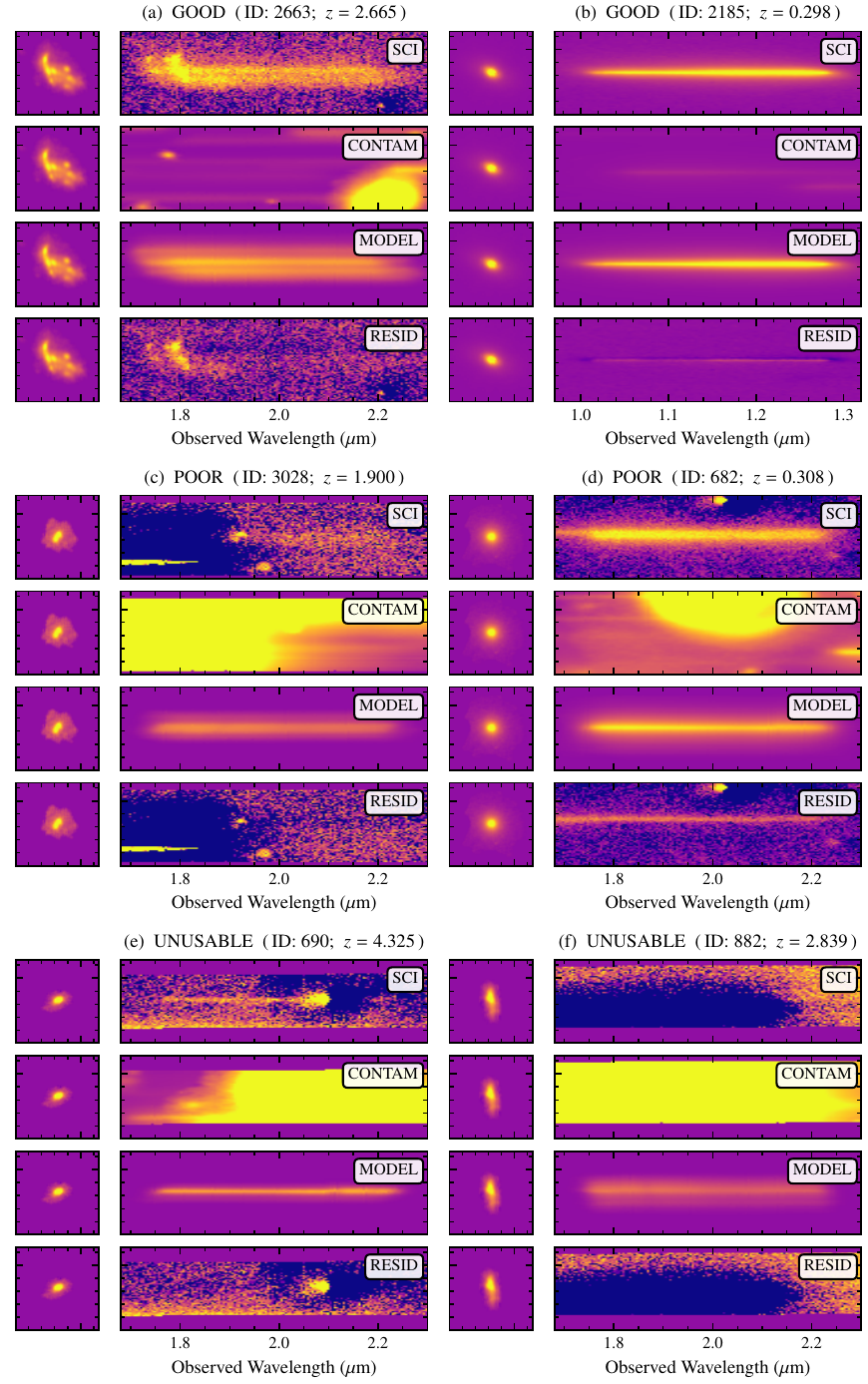}
    \caption{Examples of the various beam quality flags.
    For each object we display the direct image in the relevant filter on the left, alongside (from top to bottom), the contamination-subtracted 2D spectrum, the contaminating flux, the best-fit continuum model, and the residual after continuum subtraction.
    }
    \label{fig:beam_quality_examples}
\end{figure*}

We display in Fig.~\ref{fig:beam_quality_examples} examples of the beam quality flags, ranging from `Good' to `Unusable'.
Figures~\ref{fig:beam_quality_examples}a and \ref{fig:beam_quality_examples}b were both classified as `Good'.
Figure~\ref{fig:beam_quality_examples}a shows a high-redshift galaxy, with clear spatially resolved line emission.
Although some contamination is visible in the lower-right-hand corner, this has almost no impact on the dispersed spectrum.
Figure~\ref{fig:beam_quality_examples}b is a lower-redshift cluster member.
The low-level residual flux here is typical of such galaxies, and stems from small systematic offsets in the NIRISS wavelength calibration (see also Sect.~\ref{sec:results_redshift_validation}).

In Figs.~\ref{fig:beam_quality_examples}c and d, we show beams classified as `Poor'.
These represent data that still offer some scientific value, but should be treated with caution.
For example, in Fig.~\ref{fig:beam_quality_examples}c, a clear \Halpha\ line is visible at $\approx1.9\,\mu$m, enabling a secure redshift confirmation, despite the oversubtraction of the overlapping contamination.
Similarly for Fig.~\ref{fig:beam_quality_examples}d, the residual plot shows the unsubtracted flux from a contaminating source, which originated outside the field of view of the NIRISS direct imaging.
However, this is much fainter than the galaxy itself and is therefore likely to have had only a minor effect on the spectral extraction.

In Figs.~\ref{fig:beam_quality_examples}e and f, we display beams that were flagged as `Unusable'.
The scientific value of these data were assessed to be minimal, and they were rejected from further analysis.
Figure~\ref{fig:beam_quality_examples}e is extremely heavily contaminated, and the spectrum is dominated by an undersubtracted $0^{\rm{th}}$ order, centred around $\approx2.1\,\mu$m.
This can be distinguished from Fig.~\ref{fig:beam_quality_examples}c, as the morphology of the residual `emission' does not match the morphology of the source in the direct imaging.
Figure~\ref{fig:beam_quality_examples}f presents a simple case, where the contamination has been oversubtracted to such an extent that none of the dispersed spectrum remains.

\section{Examples: Redshift quality}\label{app:examples_redshift_quality_flag}

\begin{figure*}
    \centering
    \includegraphics[width=0.95\textwidth]{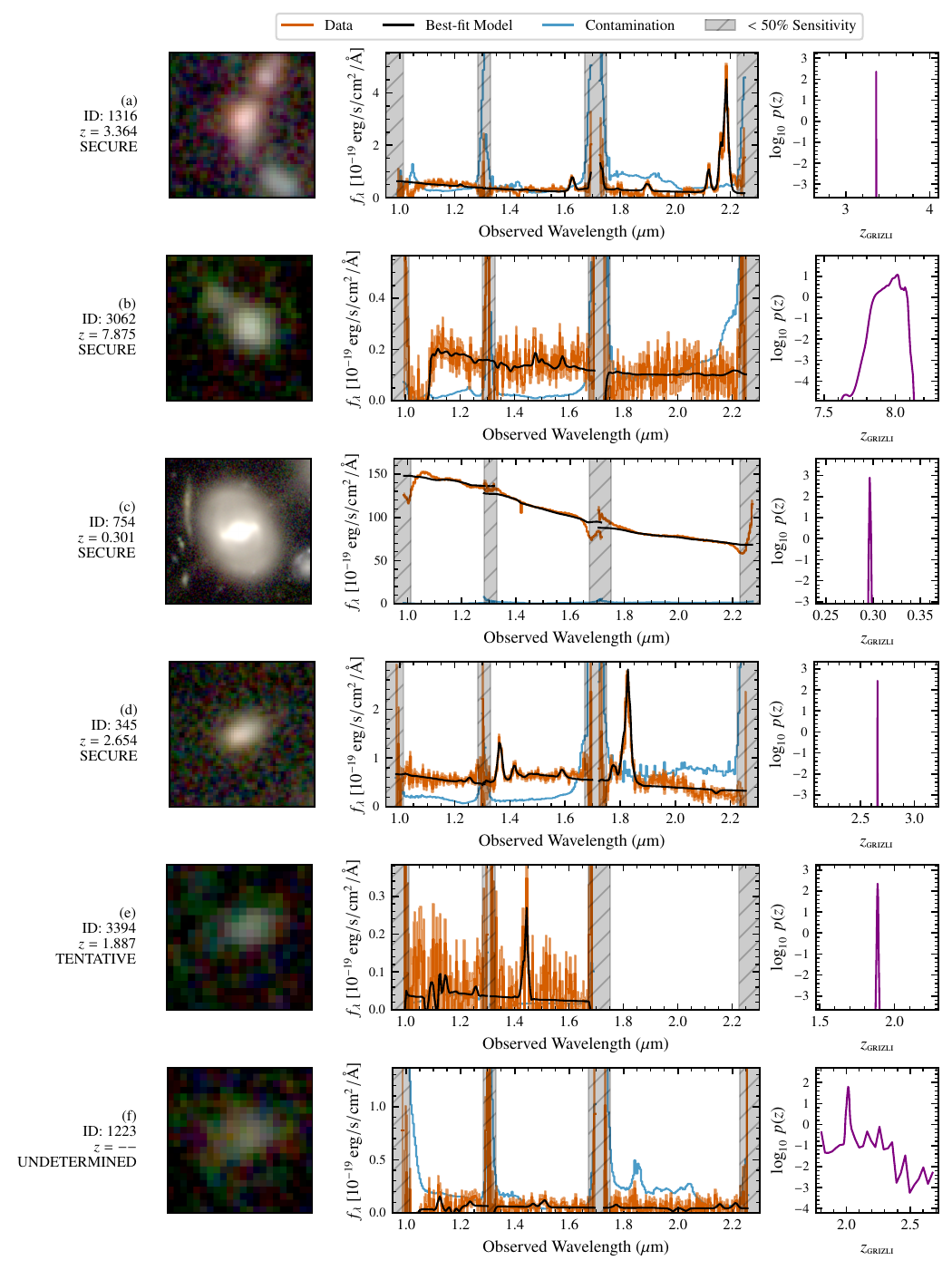}
    \caption{Examples of the various redshift quality flags.
    Left: RGB cut-outs of each object, created in the same manner as in Fig.~\ref{fig:multiple_images_cutouts}.
    Centre: Extracted 1D spectra, with the same colour scheme as in Fig.~\ref{fig:niriss_beams_example}.
    We also overlay the best-fit model spectrum for each source as a black line.
    Right: Redshift grid search, showing the (logarithmic) probability distribution as a function of redshift.
    }
    \label{fig:redshift_quality_examples}
\end{figure*}

We display in Fig.~\ref{fig:redshift_quality_examples} examples of the redshift quality flags, ranging from `Secure' to `Undetermined'.
Figs.~\ref{fig:redshift_quality_examples}a-d display galaxies with secure spectroscopic redshifts.
In Figs.~\ref{fig:redshift_quality_examples}a and d, we highlight galaxies where the redshift is secured by the detection of multiple clear emission lines, including \Oii, \Hgamma, \Hbeta, and \Oiii, giving an extremely narrow peak in the probability distribution.
We display in the same panel the contaminating flux from overlapping objects.
Regions with high contamination visibly correspond to areas where the best-fit model cannot accurately reproduce the continuum level (Fig.~\ref{fig:redshift_quality_examples}a, $\lambda\lesssim1.1\,\mu$m; Fig.~\ref{fig:redshift_quality_examples}d, $\lambda\gtrsim1.9\,\mu$m), demonstrating the limitations of slitless spectroscopic data.

Figure~\ref{fig:redshift_quality_examples}b is an example of a high-redshift galaxy, determined by the existence of a clear Lyman break.
This single feature, combined with the low continuum S/N, leads to a much broader redshift probability distribution when compared with galaxies with strong emission lines.
Similarly, Fig.~\ref{fig:redshift_quality_examples}c shows a galaxy at the cluster redshift, with the characteristic double-dip in the F115W filter (see Sect.~\ref{sec:cluster_galaxies}).
The extended nature of this object also leads to unphysical peaks and troughs in the spectrum at the edges of each filter (seen also in Fig.~\ref{fig:absorption_bands}), due to the algorithm used to extract the 1D spectrum.

Figure~\ref{fig:redshift_quality_examples}e is an example of a tentative redshift, driven by \Oiii.
At $z\sim1.9$ one would normally expect to observe \Halpha\ in the F200W filter, leading to a secure redshift, but for this particular galaxy, neither orientation was deemed usable.
Lastly, Fig.~\ref{fig:redshift_quality_examples}f gives an example of an `Undetermined' redshift.
No clear features are visible in the spectrum, and throughout most of the wavelength range, the continuum level of the source is lower than the contamination from other objects.
This is reflected in the probability distribution, which shows substantially more variation than in any of the other examples above.

\end{appendix}
\end{document}